# Ultraviolet Resonant Nanogap Antennas with Rhodium Nanocube Dimers for Enhancing Protein Intrinsic Autofluorescence

Prithu Roy,[1] Siyuan Zhu,[2] Jean-Benoît Claude,[1] Jie Liu,[2] Jérôme Wenger[1,*]

[1] Aix Marseille Univ, CNRS, Centrale Marseille, Institut Fresnel, AMUTech, 13013 Marseille, France

[2] Department of Chemistry, Duke University, Durham, NC 27708, USA

* Corresponding author: jerome.wenger@fresnel.fr

**Abstract:**
Plasmonic optical nanoantennas offer compelling solutions for enhancing light-matter interactions at the nanoscale. However, until now, their focus has been mainly limited to the visible and near-infrared regions, overlooking the immense potential of the ultraviolet (UV) range, where molecules exhibit their strongest absorption. Here, we present the realization of UV resonant nanogap antennas constructed from paired rhodium nanocubes. Rhodium emerges as a robust alternative to aluminum, offering enhanced stability in wet environments and ensuring reliable performance in the UV range. Our results showcase the nanoantenna ability to enhance the UV autofluorescence of label-free streptavidin and hemoglobin proteins. We achieve significant enhancements of the autofluorescence brightness per protein by up to 120-fold, and reach zeptoliter detection volumes enabling UV autofluorescence correlation spectroscopy (UV-FCS) at high concentrations of several tens of micromolar. We investigate the modulation of fluorescence photokinetic rates and report excellent agreement between experimental results and numerical simulations. This work expands the applicability of plasmonic nanoantennas into the deep UV range, unlocking the investigation of label-free proteins at physiological concentrations.

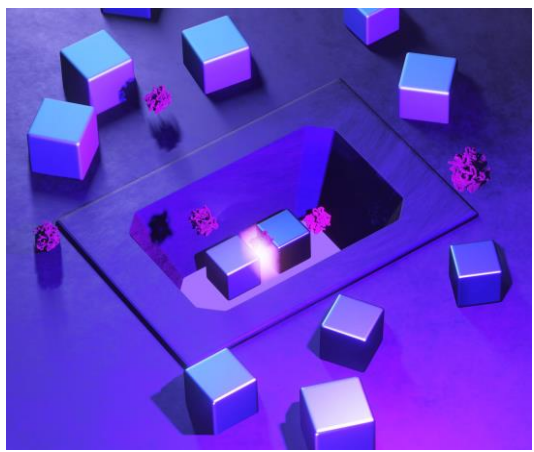

Figure for Table of Contents

**Introduction**

The interaction between light and a single fluorescent molecule is fundamentally limited by the over 100-fold size mismatch between their respective wavelengths and dimensions,[1] leading to a weak net fluorescence signal per molecule in diffraction-limited microscopes. As a result, the sensitivity and temporal resolution of single molecule fluorescence techniques, which are essential in modern biophysics and biochemistry, are also limited.[2,3] To overcome these limits, plasmonic optical nanoantennas have been introduced to manipulate light at the deeply subwavelength scale and enhance the light-matter interactions.[4,5] A broad range of optical nanoantenna designs including bowtie,[6,7] single nanorod,[8–10] nanoparticle on mirror,[11–13] nanoparticle assemblies,[14] DNA-origami dimer,[15–21] DNA-templated dimer,[22–24] metasurface,[25–27] or antenna-in-box,[28,29] has been demonstrated to significantly enhance the fluorescence brightness of single molecules, reaching impressive fluorescence enhancement factors above 1000-fold. By manipulating light at the nanoscale, the optical nanoantennas provide exquisite control over the radiation properties of a single quantum emitter, providing various possibilities to tune the directionality of the fluorescence light,[30,31] achieve ultrafast photoemission,[11,12] reduce the photobleaching rate,[32–34] control the near-field dipole-dipole energy transfer,[35,36] or trap single nano-objects.[37,38]

However, the current demonstrations and operating range of nanoantennas remain largely limited to the visible and near-infrared regions. While this range is well suited for organic fluorophores and quantum dots, extending it towards the ultraviolet (UV) region brings the key additional benefit of exploiting directly the autofluorescence of proteins without requiring any additional fluorescent label.[39–47] Over 90% of all human proteins contain tryptophan or tyrosine amino acid residues which are naturally fluorescent in the UV.[40,48] Exploiting this intrinsic UV autofluorescence signal is an appealing route to monitor single label-free proteins, releasing the need for any external fluorescence labelling.[49] The issues related to external fluorescence labelling are not only a matter of time and cost of preparation, but mainly the potential adverse effects the fluorescent marker may have on the protein conformation and/or dynamics, as documented by several reports.[50–59]

Despite the growing interest in utilizing the UV range to enhance the light-matter interaction, there have been limited reports about UV resonant optical nanoantennas.[60] Earlier works concerned mostly aluminum nanoparticle arrays to enhance Raman scattering,[61–66] and fluorescence[67–72] from dense molecular layers. Another important class of UV nanoantennas is subwavelength nanoapertures,[73–79] which can be combined with a microreflector to increase the collection efficiency.[48,80] However, all these designs are only weakly resonant, and lack the strong field confinement achieved with gap surface plasmon resonances.[1] Although numerical investigations have explored UV resonant nanoparticles[81,82] and dimer gap antennas[83–87] to achieve higher local field enhancement, their experimental demonstrations have been limited so far to Raman scattering[88] and near-field imaging.[89] The major application in enhancing the autofluorescence of label-free proteins remains unexplored. Beyond the challenging difficulty of such experiments, another limiting factor is the poor stability of aluminum plasmonics structures in aqueous environment,[90–93] especially under UV illumination.[94,95] While coating with silica or other oxide materials can promote the Al corrosion stability,[93,95] this comes at the expense of a ~10 nm-thick supplementary layer which in turn enlarges the gap size and reduces the net enhancement in the antenna hotspot.

Here we simulate, fabricate and characterize UV resonant nanogap antennas made of dimers of rhodium nanocubes to enhance the tryptophan autofluorescence from label-free proteins. Rhodium nanoantennas provide a powerful solution to the water corrosion issue associated with aluminum,[96,97] while maintaining a good plasmonic response down to the deep UV range.[81] Our fabrication approach relies on capillary assisted self-assembly of rhodium nanocubes into rectangular nanoholes milled into a quartz substrate,[98,99] leaving the nanogap region free of organic molecules for detecting diffusing proteins over a minimal luminescence background (Fig. 1a and S1). Our UV resonant nanogap antennas provide enhancement factors up to 120-fold for the autofluorescence brightness of single proteins. The UV light is concentrated into 40 zeptoliter detection volumes, which in turn enables UV autofluorescence correlation spectroscopy (UV-FCS) at concentrations exceeding 50 µM.[100,101] We also investigate the modification of different fluorescence photokinetic rates by rhodium nanoantennas and demonstrate an excellent agreement between our experiments and numerical simulations. Overall, our study expands the applicability of plasmonic nanoantennas down to the deep UV range,[4,5] broadening the capabilities to interrogate single proteins in their native state at physiological concentrations.[100–102]

**Results and Discussion**

The synthesis of the rhodium nanocubes follows the protocol published in the literature[96] based on slow injection of polyols. This approach allows a precise tunability of the nanocube size as well as a narrow size distribution. The key advantages of rhodium in this context are (i) the precise control on the nanocube size and shape allowing the tuning of the plasmonic resonance down into the UV region, (ii) the resistance to UV-induced photocorrosion largely outperforming aluminum,[94,95] and (iii) the absence of a native oxide layer to maximize the nanogap enhancement. Here, we have selected a nanocube size around 30 nm (Fig. 1b). As we discuss below, this size ensures that the plasmonic resonance of the dimer antenna occurs near 350 nm which matches with the peak autofluorescence emission of tryptophan.[73]

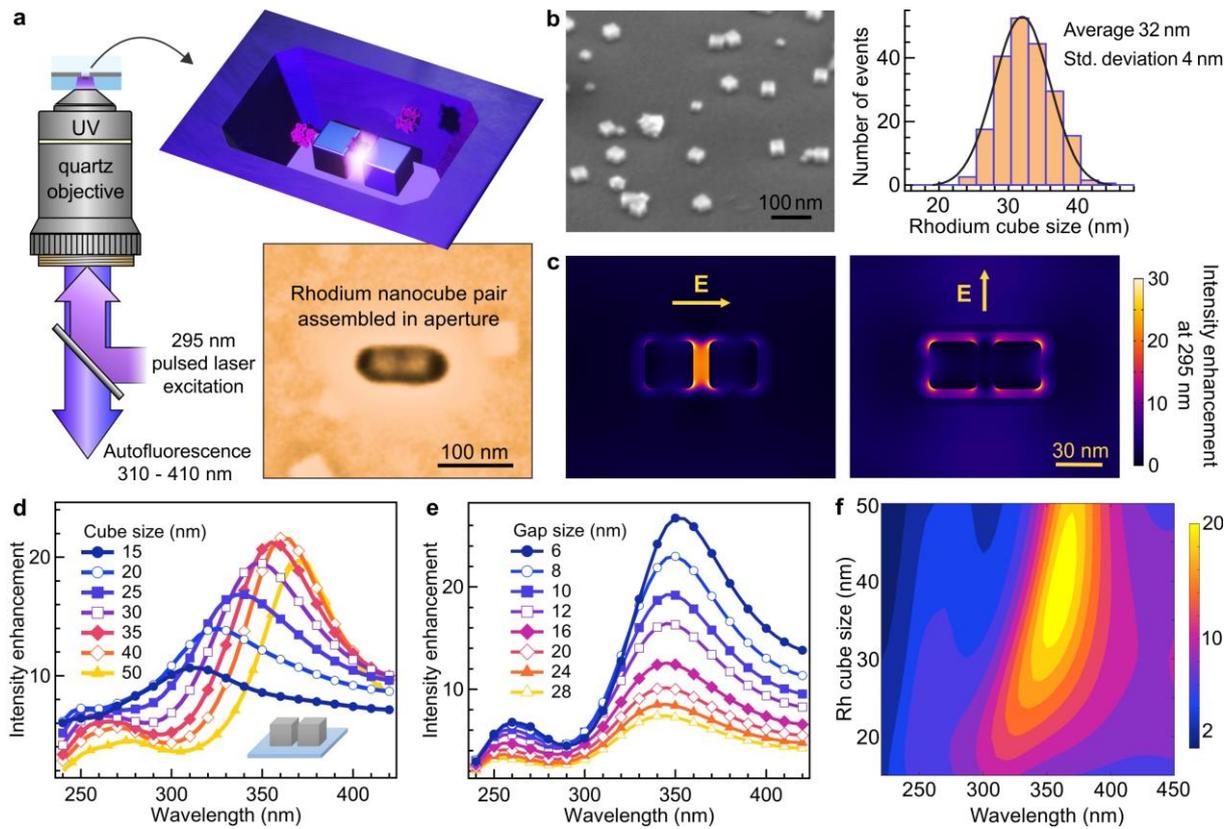

**Figure 1.** UV nanogap antenna assembled with rhodium nanocubes. (a) Scheme of the UV microscope with the resonant nanogap antenna made from two 30 nm rhodium nanocubes assembled into a 120 x 50 nm² rectangular aperture milled in an opaque aluminum film. The insert shows a scanning electron microscope (SEM) image of a nanoantenna. (b) SEM image of rhodium nanocubes dispersed on an ITO-coated coverslip viewed at 52° incidence. The size distribution correspond to the length of the rhodium nanocubes as measured by SEM, without any deconvolution or data treatment. (c) Numerical simulations of the electric field intensity enhancement at 295 nm for a UV nanogap antenna made of two 30 nm rhodium cubes separated by a 10 nm gap in a rectangular nanoaperture milled into an aluminum film. The arrows indicate the orientation of the incident electric field. The device is immersed in water. (d-f) Numerical simulations of the spectral dependence of the intensity enhancement in the center of the nanogap antenna as a function of the rhodium nanocube size and gap size. For (d) and (f) the gap size is set at 10 nm, while for (e) the cube size is 30 nm. To speed up the numerical calculations and provide design guidelines, the simulations in (d-f) consider only a pair of rhodium nanocubes on a quartz coverslip immersed in water, there is no aluminum layer here.

The fabrication of the dimer nanogap antennas relies on the capillary-assisted self-assembly applied to the rhodium nanocubes.[98,99] Focused ion beam (FIB) is used to mill rectangular nanoapertures into an aluminum-covered quartz substrate to serve as template for the nanocubes self-assembly (see Methods for complete experimental details). The 120 × 50 nm² size of the nanoaperture is chosen to accommodate only two nanocubes and leave a ~10 nm gap between them (Fig. S1 & S2). Figure 1a shows a typical scanning electron microscope (SEM) image of an assembled nanogap antenna, with more examples provided in Fig. S1 of the Supporting Information. Correlative measurements between the SEM and the UV microscope using fiducials on the sample allow to select only the antennas where a dimer of rhodium nanocubes is clearly seen (Fig. S3). Importantly in this study, Fig. S1 shows the SEM image for each nanoantenna probed in the UV microscope. Each antenna is identified with an alphanumeric code together with a specific symbol, allowing to correlate the specific geometry of the nanoantenna with its optical performance. We meticulously evaluate the gap sizes for each SEM image (Fig. S1) and consistently achieve gap dimensions ranging between 10 and 20 nm, with a median gap size of 14 nm. These results exhibit favorable comparisons with top-down fabrication methodologies, such as focused ion beam and electron-beam lithography, as illustrated in Fig. S2. However, it is important to note that the primary focus of our research is not centered on developing the nanofabrication technique. Instead, our key objective is to showcase the successful realization of ultraviolet nanogap antennas and demonstrate their performance for detecting label-free proteins. As additional advantages of our design, we benefit from the single crystallinity of the rhodium nanocubes to reduce the plasmonic losses.[1,9] The aluminum layer serves to block the direct illumination of the molecules diffusing away from the nanoantenna but still present in the confocal volume, as with the antenna-in-box design (Fig. 1a).[28,29] This method also leaves the nanogap region completely free of organic molecules which is important to reduce the residual UV luminescence background for the detection of diffusing proteins.

Numerical simulations based on the finite element method confirm the excitation of resonant nanogap modes when the excitation polarization is set parallel to the dimer main axis. Figure 1c shows the intensity maps for 295 nm excitation, which was used in our experiments on proteins as this wavelength gives a slightly better signal to background ratio than the 266 nm laser line. The intensity maps for 266 nm (p-terphenyl excitation) and 350 nm (peak autofluorescence emission) are shown in the Supporting Information Fig. S4 and S5 respectively, with an intensity profile similar to the one found for 295 nm (Fig. 1c). Even in cases of significant misalignment of the nanocubes, substantial optical confinement and intensity enhancement are still predicted by numerical simulations (Fig. S6).

Increasing the size of the nanocube leads to a red-shift of the plasmonic resonance (Fig. 1d). Reducing the gap size increases the intensity enhancement in the nanogap region and also leads to a red-shift of the resonance (Fig. 1e). Altogether, these features demonstrate the occurrence of plasmonic nanogap resonances in rhodium dimer antennas.[4,5] Using a parametric study as a function of cube size, gap size and resonance wavelength (Fig. 1f and S7), we select nanocube sizes around 30 nm for the autofluorescence enhancement experiments. This leads to a plasmonic resonance slightly blue-shifted respective to the peak protein emission wavelength at 350 nm (Fig. S8), as this condition has been proven to yield the best brightness enhancement factors.[103]

We use fluorescence correlation spectroscopy (FCS) and time-correlated single photon counting (TCSPC) experiments to assess the optical performance of the nanoantennas and their ability to enhance the UV autofluorescence of diffusing label-free proteins. The comparison between experiments performed with the excitation laser polarization set parallel and perpendicular to the main antenna axis demonstrates the contribution of the nanogap enhancement. Figure 2 summarizes the results found with label-free streptavidin at 50 µM concentration. A higher intensity is obtained when the excitation polarization is set parallel to the gap (Fig. 2a). We have checked that the excitation and detection on our microscope is not polarization sensitive, so that the polarization dependence can be directly linked with the enhanced autofluorescence signal stemming from the nanoantenna gap region. However, relying solely on the total intensity averaged across the entire antenna volume is inadequate for estimating the brightness enhancement per molecule. This is because the total intensity comprises the product of brightness and the number of molecules. To overcome this challenge, we employ FCS as a powerful technique to independently determine both the number of molecules contributing to the signal and their individual autofluorescence brightness per emitter.[14,28,29] In addition, FCS is supplemented with time-correlated single photon counting (TCSPC) to estimate the fluorescence lifetime.

Looking at the raw data, we readily observe that the FCS curve has a higher correlation amplitude with parallel excitation polarization than with perpendicular orientation (Fig. 2b, Tab. S1), while the autofluorescence lifetime is reduced when the excitation polarization is turned from perpendicular to parallel (Fig. 2c, Tab. S2). The nanoantenna significantly reduces the autofluorescence lifetime from 1.5 ns for the confocal reference to 0.47 ns for the antenna with parallel orientation, demonstrating a higher local density of optical states (LDOS) in the nanoantenna hotspot and Purcell effect on label-free proteins.[1] All these raw observations highlight the contribution of the nanogap hotspot and its effect to enhance the UV autofluorescence.

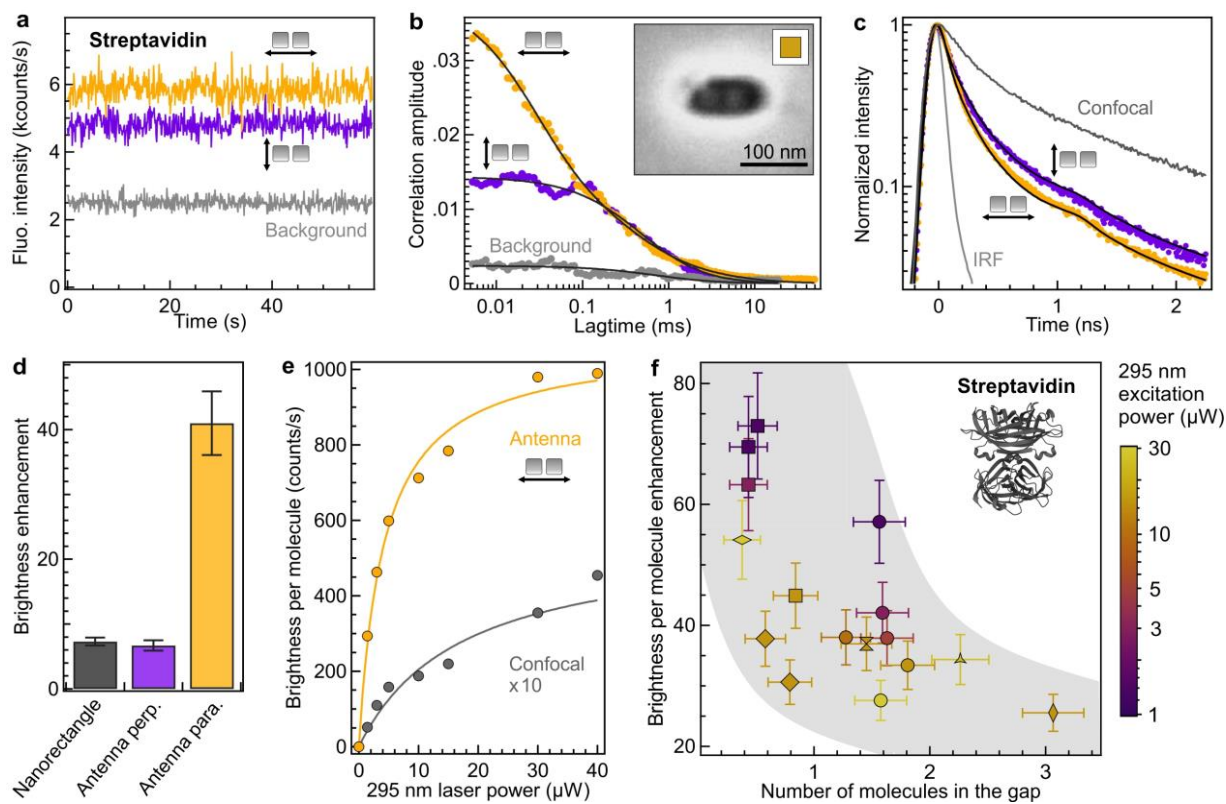

**Figure 2.** UV autofluorescence from label-free streptavidin proteins enhanced with a nanogap antenna. (a) Autofluorescence intensity time traces (binning time 100 ms) for a rhodium nanogap antenna with the excitation polarization set parallel (yellow) or perpendicular (purple) to the dimer antenna's main axis. The antenna is covered with a 50 µM solution of diffusing label-free streptavidin proteins. The gray trace shows the background intensity level in absence of the protein (the antenna is covered with the buffer solution). The 295 nm excitation power used here is 15 µW. (b) FCS correlation functions corresponding to the traces in (a). Dots are experimental data, lines are numerical fits. The insert SEM image shows the dimer antenna used for this experiment (the antenna reference is R5s1p5e2-1 with a gap size of 10 nm as measured by SEM, see Fig. S1). The data corresponding to this antenna appear as a square marker in the scatter plot (f). (c) Normalized time-resolved decay traces corresponding to the experimental data in (a) and to the confocal reference (dark gray). IRF stands for the instrument response function. (d) Comparison of the enhancement factors for the fluorescence brightness per molecule in the empty nanorectangle (without rhodium nanocubes, see Fig. S9) and the rhodium nanoantenna with parallel and perpendicular excitation polarizations. (e) Excitation power dependence of the brightness per molecule measured in the nanogap antenna (yellow markers) and in the confocal reference (gray). The line is a fit with a saturation model.[28] (f) Scatter plot of the fluorescence brightness enhancement for streptavidin proteins as a function of the number of detected molecules in the gap antenna. Different markers indicate different nanogap antennas, whose SEM images are shown in Fig. S1. The color codes for the excitation power, and the shaded area is a guide to the eyes.

To quantify the brightness enhancement with the nanoantenna, we use UV-FCS to measure the average number of molecules $N^*$ present inside the nanogap and their autofluorescence brightness per molecule $Q^*$ (see Methods). This general FCS approach has been validated previously for plasmonic antennas in the visible and organic fluorescent dyes.[14,28,29,35] For the nanoantenna with parallel excitation, we find a brightness enhancement of 41 ± 5 fold for label-free streptavidin (Fig. 2d). This performance is clearly above the enhancement found with perpendicular orientation (6.7 ± 0.8) or the empty nanoaperture in the absence of rhodium antenna (7.3 ± 0.6, see Fig. S9). We also perform experiments on single rhodium nanocubes (Fig. S10) which yield similar enhancement values as the empty nanoaperture, confirming the specific optical response from the nanogap with parallel excitation. Importantly, the enhancement factor found with the nanoantenna and parallel polarization significantly outperforms the gain obtained earlier with nanoaperture-based designs (we obtained 4-fold enhancement[78] for a bare nanoaperture without microreflector and 15-fold with the so-called horn antenna[80] combining a nanoaperture and a microreflector). This superior performance is achieved by combining UV plasmonic resonant gap modes yielding intense electromagnetic enhancements together with the corrosion-resistance and single-crystalline nature of the rhodium nanocubes.

The observation of saturation of the autofluorescence brightness (Fig. 2e) is a supplementary control to show that the signal stems from the protein autofluorescence and is not related to some laser backscattering or Raman scattering. Streptavidin autofluorescence brightness up to 1000 photons/s/molecule are reached, which is a key element in maximizing the signal to noise ratio in UV-FCS.[48] Deep UV nanoantennas offers a transformative opportunity to substantially amplify the autofluorescence signal from individual label-free proteins and thus render previously undetectable signals easily discernible. Leveraging UV-FCS experiments unlocks powerful perspectives to assess local concentrations, mobilities, brightness, and stoichiometries of label-free proteins.

For FIB[28] as well as for electron-beam lithography,[29] some variability in the nanoantenna gap size (Fig. S2) inevitably leads to a dispersion of the nanoantenna performance. We assess this effect for our UV antennas with Fig. 2f displaying the brightness enhancement as a function of the number of gap molecules $N^*$. Importantly here, each data point in Fig. 2f can be assigned to the specific SEM image of the antenna. As found earlier for visible antennas and fluorescent dyes,[29] there is a correlation between the brightness enhancement and the number of detected molecules in the gap, with antennas having a narrower gap tend to give higher brightness enhancement and lower number of molecules. We do monitor a similar feature here for UV antennas and label-free proteins. We also find a clear correlation between the volume measured with FCS and the gap size obtained from the SEM images (Fig. S11). This allows us to retrieve the evolution of the brightness enhancement as a function of the gap size (Fig. S12), where we find that the antennas with the smallest gaps provide the highest brightness enhancements.

With an average number of molecules in the nanogap of 1.3 ± 0.7 for a 50 µM concentration, the nanoantenna detection volume thus corresponds to 40 ± 20 zeptoliter (1 zL = $10^{-21}$ L = 1000 nm$^3$), 25000 fold below the femtoliter confocal detection volume. For comparison, experiments on gold nanoantennas with 12 nm gaps led to detection volumes of 100 zL, while dimers of spherical 80 nm gold nanoparticles gave volumes of 70 zL.[14] Our correlative UV-FCS and SEM measurements clearly underscore the FCS volume dependence with the gap size (Fig. S11a) with the smallest 10 nm gaps yielding volumes down to 15 zL while the largest 26 nm gaps providing volumes around 100 zL.

The UV-FCS experiments are repeated with p-terphenyl (Fig. 3) and hemoglobin (Fig. S13) to probe a wider range of conditions and initial quantum yield of the emitters. P-terphenyl is a UV fluorescent dye with 93% quantum efficiency,[73] while our comparative spectroscopy experiments estimate the average quantum yield of hemoglobin to be around 0.5% (Fig. S8). The characteristic polarization-dependent signature of the dimer gap antenna is observed again for both p-terphenyl (Fig. 3a-c) and hemoglobin (Fig. S13). Experiments with p-terphenyl are further used to control that the diffusion time across the nanoantenna scales with the solution viscosity. While replacing cyclohexane by a 60:40 (v/v) glycerol:ethanol mixture, confocal experiments calibrate that the solution viscosity increases by 12.5-fold. Our nanoantenna data (Fig. 3b) retrieve a similar increase of 11.4-fold of the diffusion time which we directly relate to the increase in the solution viscosity.

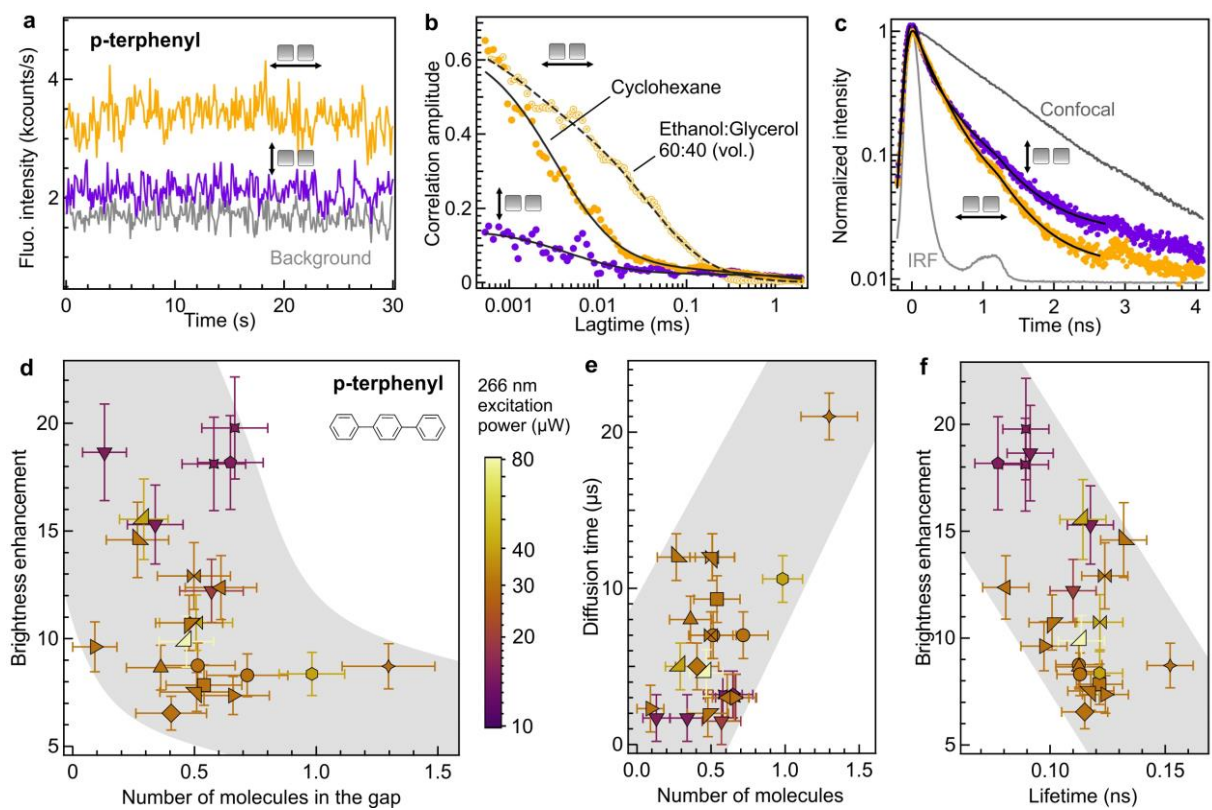

**Figure 3.** UV fluorescence enhancement of p-terphenyl with rhodium nanogap antennas. (a) Fluorescence intensity time traces recorded on a 10 µM solution of p-terphenyl in cyclohexane on a nanoantenna with the excitation polarization set parallel (yellow) or perpendicular (purple) to the dimer antenna's main axis. The binning time is 100 ms. The gray trace shows the background intensity level in absence of p-terphenyl. The 266 nm excitation power used here is 40 µW. The data in a-c correspond to the antenna reference number R10s1p2d5-8 for which a 11 nm gap size was inferred from the SEM image (Fig. S1). (b) FCS correlation functions corresponding to the traces in (a) and when p-terphenyl molecules are diluted into a 60:40 glycerol:ethanol mixture to increase the viscosity. Dots are experimental data, lines are numerical fits. (c) Normalized time-resolved decay traces corresponding to the experimental data in (a) and to the confocal reference (dark gray). (d) Scatter plot of the fluorescence brightness enhancement for p-terphenyl as a function of the number of molecules detected in the gap antenna. The various markers indicate different nanoantennas, whose SEM images are shown in Fig. S1. The color codes for the excitation power. Among the different experiments, the number of molecules have been scaled to correspond to a 30 µM concentration of p-terphenyl. (e) Scatter plot of the FCS diffusion time as a function of the number of molecules detected in the nanogap. (f) Scatter plot of the fluorescence brightness enhancement as a function of the fluorescence lifetime. Throughout (d-f), the shaded areas are guides to the eyes.

Looking at the statistics from 27 antennas, we note that smaller gap volumes lead to higher brightness enhancement factors (Fig. 3d and S12) and shorter diffusion times (Fig. 3e). We relate both features to a better confinement of light into nanoscale dimensions, as demonstrated by the SEM gap sizes (Fig. S1 and S11). The detection volume inferred from UV-FCS on p-terphenyl is 30 ± 15 zeptoliter, which stands in good agreement with the streptavidin data. Our data also display the interesting trend that the brightness enhancement scales inversely with the fluorescence lifetime (Fig. 3f). This indicates that in the range of conditions probed here, the antennas with smaller gaps lead to higher molecular brightness and higher LDOS (shorter lifetime), as expected for resonant plasmonic nanoantennas.

To understand better the physics behind the UV fluorescence enhancement and assess the influence of plasmonic losses, we perform numerical simulations and estimate the antenna's influence on the radiative, nonradiative and total decay rate constants (Fig. 4a-c, S14 and S15). Nanoantennas made of 30 nm rhodium cubes have a dipolar resonance around 350 and a quadrupolar resonance around 260 nm. The quadrupolar resonance is essentially nonradiative. This effect, together with the increased intrinsic losses of rhodium below 300 nm, explains the increase of the nonradiative rate enhancement and the drop of the antenna efficiency below 300 nm (Fig. S14). The different decay rate constants critically depend on the gap size (Fig. 4c), with the smallest gap sizes below 10 nm being dominated by nonradiative losses.

With the knowledge of the excitation intensity gain (Fig. 1c) and the antenna's influence on the various photokinetic rates, we can infer the net fluorescence brightness enhancement as a function of the emitter's initial quantum yield,[17,29] and compare with our experimental results. Emitters with lower quantum yields give higher apparent brightness enhancement factors (Fig. 4d) as a maximum benefit can be taken from the nanoantenna's ability to enhance the radiative rate.[17,29] We compare our experimental results with the numerical predictions in Fig. 4e. Within the experimental uncertainties, the enhancement values found for the different molecules agree well with the theoretical predictions, confirming the validity of our approach. Experimentally, the highest brightness enhancement of at 120-fold is obtained with hemoglobin, which has the lowest 0.5% quantum yield in solution. The simulations predict even higher enhancement factors above 400-fold, yet in the case of a dipolar source perfectly aligned with the nanoantenna (Fig. S16).

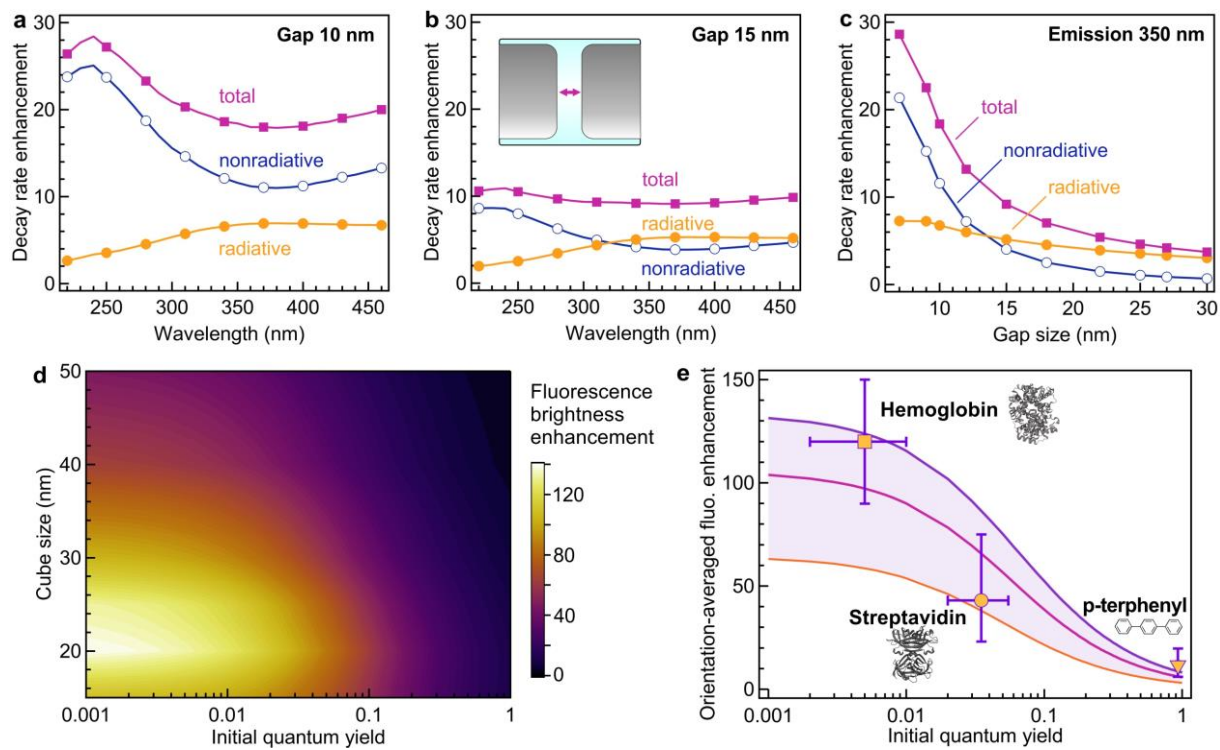

**Figure 4.** Rhodium UV nanogap antennas to enhance the photokinetic rates. (a,b) Numerical simulations of the enhancement of the decay rate constants as a function of the emission wavelength for a perfect dipole emitter with parallel orientation located in the center of the gap between two rhodium nanocubes in water. The rhodium cube size is constant at 30 nm. The gap size is 10 nm in (a) and 15 nm in (b). All rates are normalized respective to the dipole radiative rate in free space. (c) Evolution of the decay rate enhancement factors as a function of the gap size, for an emission wavelength of 350 nm and a cube size of 30 nm. (d) Simulations of the fluorescence brightness enhancement as a function of the rhodium nanocube size and the emitter's initial quantum yield in free space. The gap size is kept constant at 10 nm. The excitation wavelength is 295 nm and the emission is 350 nm. The emission is averaged over the three orientation directions. (e) Comparison of the simulated (lines) and experimental (markers) fluorescence brightness enhancement factors as a function of the quantum yield in homogeneous solution for the different emitters used in this work. The emission is averaged over the three orientation directions. From top to bottom, the three lines represent rhodium cube sizes of 25, 30 and 40 nm respectively, with a gap size set to 10 nm. The protein structures in gray have been made using Mol* viewer.[104]

Combining all the experimental results on the brightness enhancement and the fluorescence lifetime reduction, we can compute back all the different decay rate constants (Supporting Information Tab. S3). For p-terphenyl, streptavidin and hemoglobin, and despite the large difference in their initial quantum yields, we find consistent excitation gains $\eta_{exc} = 15.5 \pm 3.8$ and radiative gains $\eta_{\Gamma rad} = 10.8 \pm 2.6$ in good agreement with numerical simulations considering the experimental values are orientation-averaged and position-averaged inside the nanogap. The loss decay rate constant into the metal is also a preserved feature among our different experiments, with $\Gamma_{loss}^* = 1.25 \pm 0.3$ ns$^{-1}$. Comparing with nanogap antennas in the red spectral range with comparable gap sizes, a loss decay rate constant of 0.5 ns$^{-1}$ can be found for gold,[16,23] while aluminum and silicon yield typically 2 ns$^{-1}$ and 4 ns$^{-1}$ respectively.[35,105] The nonradiative losses associated to rhodium in the UV appear thus quite comparable to other materials in the visible range.

In the future, aluminum nanocubes[106] and nanocrystals[93] could allow to reach lower losses than rhodium, provided their water corrosion issue can be circumvented. We have performed numerical simulations to assess the UV performance of optimized aluminum nanoantennas and compare them with the rhodium antennas discussed in this work. Figure S17 in the Supporting Information summarizes our main results. We find that pure aluminum outperforms rhodium by approximately 50%, aligning with our initial expectations. However, the introduction of a supplementary oxide layer to safeguard aluminum against UV photocorrosion leads to a notable drop in performance, with the enhancement being 3 times lower for protected aluminum as compared to pure rhodium. These compelling results further underscore the interest for rhodium in UV plasmonics applications and the need for specific care while designing protective measures for the nanoantennas.

**Conclusions**

In conclusion, our work provides experimental and numerical evidence for the successful implementation of rhodium nanogap antennas with plasmonic resonances extending deep into the ultraviolet region. By harnessing the combination of intense electric field enhancement and photokinetic rates alteration, our antenna design achieves brightness enhancement factors up to 120-fold, together with detection volumes in the zeptoliter range and sub-nanosecond autofluorescence lifetime. Notably, correlative SEM and UV-FCS measurements demonstrate that the nanogap mode plays a pivotal role, as evident from polarization-dependent measurements and the interdependence observed among brightness, lifetime, and detection volume. Thanks to the intense nanogap enhancement, the plasmonic resonant nature of the gap mode and the single crystallinity and smooth surface of the rhodium nanocubes, the optical performance of our antennas significantly outperforms previous nanoparticle- or nanoaperture-based devices.

Enhancing the autofluorescence of label-free proteins stands as a major application and driving motivation for UV plasmonics. To showcase this capability, we present the successful enhancement of autofluorescence signals from streptavidin and hemoglobin proteins. While the autofluorescence quantum yield of tryptophan in most proteins is typically on the order of a few percent,[47] there is a compelling interest in utilizing UV nanoantennas to significantly amplify the autofluorescence signal from single proteins, rendering it easily detectable. Leveraging UV-FCS experiments, we unlock powerful perspectives for local measurements of concentration, mobility, brightness, and stoichiometry of label-free proteins.[100–102]

Altogether, our work significantly advances the field of nanotechnology and biosensing by demonstrating the success of ultraviolet nanogap antennas for label-free protein detection. Extending the practical application of plasmonic nanoantennas into the deep UV range broadens the capabilities to investigate individual proteins in their native state under physiological concentrations.[100–102] The robustness of the achieved gap sizes further enhances the practical applicability of our approach, positioning it as a promising technique in this domain. Beyond label-free protein autofluorescence detection, resonant UV nanoantennas are highly relevant to advance several other plasmonic applications, including resonant Raman spectroscopy,[61,79] circular dichroism spectroscopy,[107] photodetectors,[108] and photocatalysis.[109]

**Materials and Methods**

*Rhodium nanocube synthesis.*

Rhodium nanocubes were synthesized using a seed-mediated method reported earlier.[96] First, rhodium seed solution was prepared: 0.45 mmol KBr (ACROS, reagent ACS) was dissolved in 2 mL ethylene glycol (J.T. Baker, 99.0%) in a 20 mL scintillation vial. The vial was put in the oil bath for 40 min at 160°C. Subsequently, 0.045 mmol $RhCl_3 \cdot xH_2O$ (Aldrich, 98%) and 0.225 mmol PVP (Aldrich, mw=55000) were dissolved in 2 mL ethylene glycol separately. A two-channel syringe pump was used to pump these two solutions into the vial at a speed of 1 mL/h. The solution was aged for 10 min at 160°C before cooling to room temperature as the seed solution. 0.4 mL of the prepared rhodium seed solution was then mixed with 1.6 mL ethylene glycol for a total 2 mL in another 20 mL scintillation vial. The vial was put in the oil bath for 40 min at 160°C. Two additional solutions produced by dissolving 0.045 mL $RhCl_3 \cdot xH_2O$ in 2 mL ethylene glycol and 0.225 mmol PVP (Aldrich, mw=55000) plus 0.45 mmol KBr were in another 2 mL ethylene glycol. These two solutions were pumped into the heated vial at a speed of 1 mL/h. After all solutions were added, the mixture was cooled to room temperature and rhodium nanocubes were collected after centrifugation and washes with water/acetone for several times.

*Nanoantenna fabrication.* Arrays of 120 nm by 50 nm rectangular nanoapertures together with fiducial marks are milled by focused ion beam (FIB) on a UV-transparent quartz coverslip substrate covered with 100 nm thick aluminum layer. FIB milling is performed on a FEI DB235 Strata with 30 kV acceleration voltage and 10 pA gallium ion current. The aluminum nanoapertures are covered by a 10 nm-thick silica layer deposited with plasma-enhanced chemical vapor protection (Oxford Instruments PlasmaPro NGP80) in order to protect the aluminum layer against UV-induced photocorrosion.[94,95] For the deposition of rhodium nanocubes and their self-assembly into nanogap antennas, 2 mM sodium dodecyl sulfate (SDS) and 1% Tween20 are added to the 100 µL rhodium solution. This solution is then left for 15 minutes in an ultrasonic bath to ensure all nanoparticles are well dispersed. Droplets of 4 µL are then deposited on the aluminum nanorectangle sample and left to evaporate. Different movements of the droplet respective to the sample have been tried to benefit from capillary-assisted self-assembly, but the simple horizontal evaporation of the rhodium nanocube droplet gave the best results in our case. As we are illuminating from below the quartz substrate, the presence of extra nanoparticles on top of the aluminum film has no effect on our measurements. The nanoantennas are then imaged with a scanning electron microscope (electron beam of the FEI DB235 Strata). The position of the antennas containing two rhodium nanocubes is noted and used later to find the same antennas in the UV microscope. The rhodium antennas are remarkably stable, the sample can be rinsed and dried several times without disturbing the antenna geometry.

*Fluorescent samples.* p-terphenyl, *Streptomyces avidinii* streptavidin and human hemoglobin are purchased from Sigma-Aldrich in powder form (see complete details in Table S4). P-terphenyl is dissolved in cyclohexane, while the proteins are dissolved in a 25 mM Hepes, 300 mM NaCl, 0.1 v/v% Tween20, 1 mM DTT, and 1 mM EDTA 1 mM buffer solution at pH 6.9. The solutions are centrifuged for 12 min at 142,000 g (Airfuge 20 psi), the supernatants are stored at -20°C and further used for the experiments. The concentrations are assessed with a Tecan Spark 10M fluorometer. Prior to the UV measurements, the oxygen dissolved in the solution is removed by bubbling the buffer with argon for 5 minutes, and 10 mM of mercaptoethylamine MEA is added to improve the photostability.

*UV microscope.* For protein experiments we use a 295 nm laser (Picoquant Vis-UV-295-590) while for p-terphenyl we use a 266 nm laser (Picoquant LDH-P-FA-266). Both lasers are pulsed with 70 ps duration and 80 MHz repetition rate. The laser beams are spatially filtered by a 50 µm pinhole to achieve quasi-Gaussian beam profiles. The UV microscope objective is LOMO 58x 0.8 NA with water immersion. The nanoantenna sample is scanned by a 3-axis piezoelectric stage (Physik Instrumente P-517.3CD). The fluorescence light is collected by the same microscope objective, separated from the excitation laser beam by a dichroic mirror (Semrock FF310-Di01-25-D) and focused onto a 50 µm pinhole by a quartz lens with 200 mm focal length (Thorlabs ACA254-200-UV). Emission filters (Semrock FF01-300/LP-25 and FF01-375/110-25) are placed before the photomultiplier tube (Picoquant PMA 175) whose output is connected to a time-correlated single photon counting TCSPC module (Picoquant Picoharp 300 with time-tagged time-resolved mode). The full width at half maximum of the instrument response function is 140 ps defining the temporal resolution of our UV microscope.

*FCS analysis.* The fluorescence time traces data are computed with Symphotime 64 (Picoquant) and fitted with Igor Pro 7 (Wavemetrics). The FCS analysis builds on our previous works on nanoantennas in the visible range.[14,28,29,35] For the rhodium nanoantennas with parallel excitation, the FCS correlations are fitted with a three-species model:[110]

$$G(\tau) = \sum_i \rho_i \left(1 + \frac{\tau}{\tau_i}\right)^{-3/2} \quad (1)$$

where $\rho_i$ and $\tau_i$ are the amplitude and diffusion time of each species. Here for further simplification we have assumed that the aspect ratio of the axial to transversal dimensions of the detection volume is equal to 1 following our previous works. The rationale behind these three species model is that the first fast-diffusing species accounts for the molecules inside the nanogap, the second intermediate diffusing species account for the molecules present inside the nanorectangle but diffusing away from the nanogap hotspot while the third slowly diffusing term is introduced to account for some residual correlation stemming from the background.[48] For p-terphenyl, owing to the fast diffusion time and the high quantum yield of the dye, we find that a two-species model is sufficient to fit the FCS function. For the antennas with perpendicular excitation polarization, we always use only a two-species model as the nanogap contribution is absent in this case. Typical fit results are detailed in the Supporting Information Table S1 for the three different target molecules probed here.

Building on our earlier works on plasmonic antennas in the visible,[14,28,29,35] we use the following notations in our analysis of the antenna's performance: the average number of molecules present inside the nanogap is $N^*$ with a brightness per molecule $Q^*$. The number of molecules diffusing outside the nanogap (but still contributing to the total detected fluorescence and hence to the FCS amplitude) is $N^0$ with a brightness per molecule $Q^0$. The total fluorescence intensity is $F$ and $B$ is the background intensity recorded on the same nanoantenna in the absence of the target protein. The general FCS formalism in presence of multiple species[110] can be inversed to express the number of molecules inside the nanogap $N^*$ and their brightness per molecule $Q^*$:

$$N^* = \frac{(F - B - N^0 Q^0)^2}{F^2(\rho_1 + \rho_2) - N^0(Q^0)^2} \quad (2)$$

$$Q^* = \frac{F^2(\rho_1 + \rho_2) - N^0 (Q^0)^2}{F - B - N^0 Q^0} \qquad (3)$$

For the values of the parameters $N^0$ and $Q^0$ for the molecules diffusing outside the nanogap, we use the FCS measurements when the excitation polarization is set perpendicular to the dimer axis. We have checked the validity of these values by comparing with the results obtained with an empty nanorectangle without rhodium nanocubes. The enhancement factor for the fluorescence brightness per molecule is then computed as the ratio between $Q^*$ and the reference brightness per molecule value $Q_{ref}$ obtained from confocal FCS measurements. In the case of hemoglobin, instead of FCS experiments, we use the known protein concentration and the calibrated confocal volume of 1.8 fL to estimate the number of molecules and their brightness in the diffraction-limited confocal setup.

*Lifetime analysis.* The fluorescence decay histograms are computed and analyzed with Symphotime 64 (Picoquant). We use an iterative reconvolution fit taking into account the measured instrument response function (IRF). The decay histograms in the nanoantenna are fitted with a three-component exponential model. To ease the comparison between the parallel and perpendicular polarizations, we use the same characteristic lifetimes for both polarizations, and compute the intensity-averaged lifetime as the final readout. All the fit parameters are summarized in Table S2.

*Numerical simulations.* The electric field distributions are computed using the wave optics module of COMSOL Multiphysics v5.5, relying on the finite element method. The reflections from the boundaries of the simulation domain are suppressed using scattering boundary conditions. In our design, we round the edges of the rhodium nanocubes with a 5 nm radius of curvature to avoid any spurious effects from sharp edges. The refractive index parameters are taken from predefined libraries of COMSOL Multiphysics. To reproduce the experimental conditions, all the simulations are performed with the antennas immersed in a water environment on top of the quartz coverslip. To optimize the antenna design and explore a broad range of parameters, we use a 2D model which was checked to give comparable results to a full 3D simulation. We use a tetrahedral user-defined mesh, with mesh size ranging from 0.01 nm to 10 nm for the 2D model and 01 nm to 10 nm for the 3D model. To calculate the excitation intensity enhancement spectra, the rhodium dimer is excited with a plane wave stemming from the quartz substrate with wavelength ranging from 220 nm to 450 nm. To calculate the radiative rates enhancements, we define two monitors surrounding the source dipole one few wavelength far-from dipole to calculate the radiative power and one only few nanometer away from the source to calculate the total dissipated power. The antenna influence is determined by comparing with a similar dipolar source near a quartz substrate in water medium in the absence of the rhodium nanocubes. The convergence is checked by generating the error over iteration chart in-built in COMSOL.

**Supporting Information**

Correlative SEM images, Comparison with other nanofabrication methods, Overview of several antennas, Intensity enhancement at 266 nm and 350 nm, S5. Influence of the nanocube tilt, Spectral and size dependence, Autofluorescence emission spectra, Control FCS in the absence of rhodium nanoantenna, Control FCS with a single rhodium nanocube, Correlation between FCS volume and gap size, Brightness enhancement as a function of SEM gap size, Nano-antenna enhanced autofluorescence of hemoglobin, Fitting parameters results, Numerical simulations of decay rates enhancement, Photokinetic rates, Comparison with aluminum nanogap antennas, Protein information

**Acknowledgements**

This project has received funding from the European Research Council (ERC) under the European Union's Horizon 2020 research and innovation programme (grant agreement No 723241). Work at Duke is in part supported by NSF (CHE-1954838).

**Conflict of Interest**

The authors declare no conflict of interest.

**Data Availability Statement**

The data that support the findings of this study data are available from the corresponding author upon request.

**Supporting Information for**

**Ultraviolet Resonant Nanogap Antennas with Rhodium Nanocube Dimers for Enhancing Protein Intrinsic Autofluorescence**

Prithu Roy,[1] Siyuan Zhu,[2] Jean-Benoît Claude,[1] Jie Liu,[2] Jérôme Wenger[1,*]

[1] Aix Marseille Univ, CNRS, Centrale Marseille, Institut Fresnel, AMUTech, 13013 Marseille, France

[2] Department of Chemistry, Duke University, Durham, NC 27708, USA

* Corresponding author: jerome.wenger@fresnel.fr

**Contents:**

## S1. Correlative SEM images of the rhodium nanogap antennas

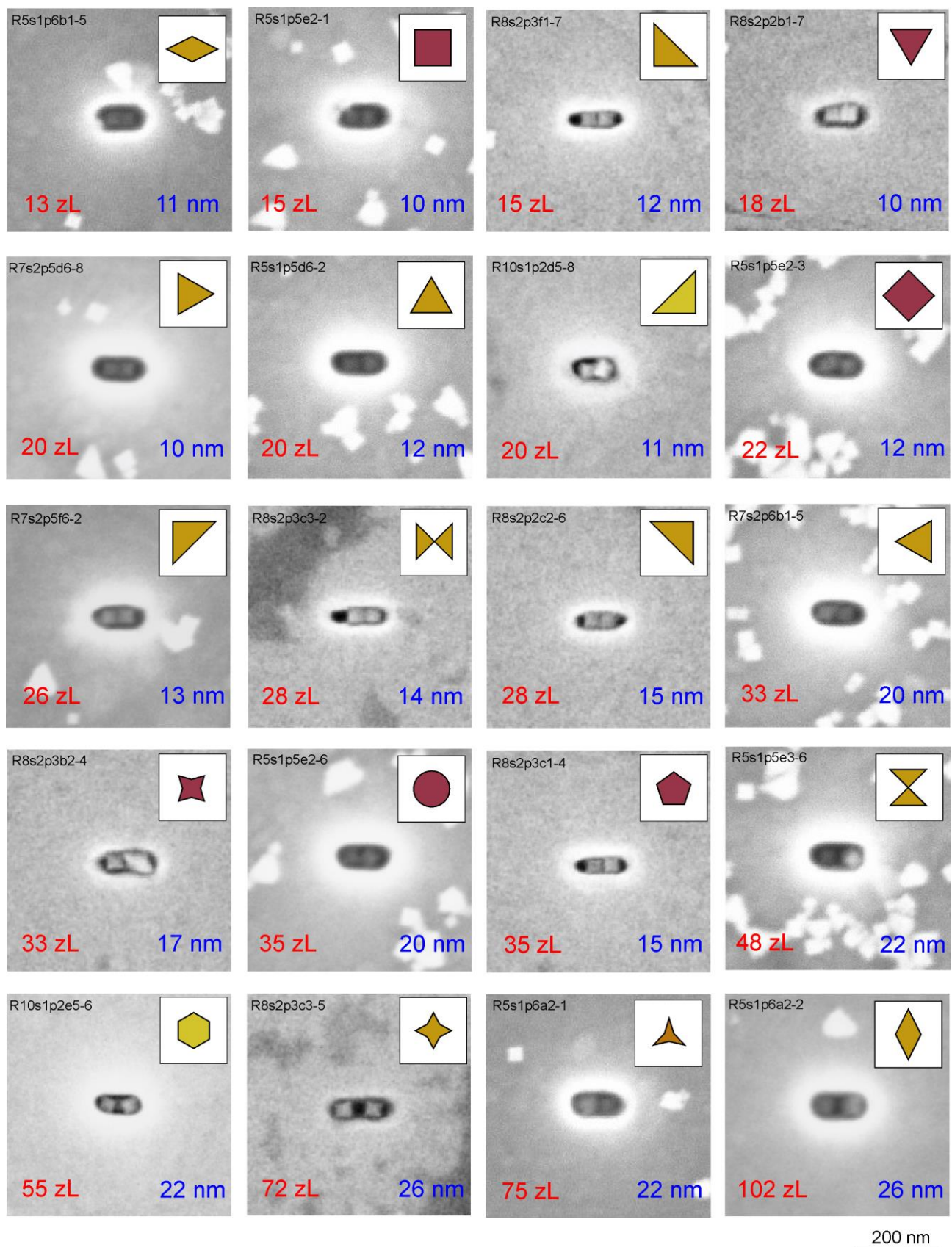

Figure caption appears on next page

**Figure S1.** Scanning electron microscopy images of the rhodium nanogap antennas used in the experiments corresponding to the data displayed in Fig. 2 & 3. The color symbol in the top right corner of each SEM image corresponds to the symbol used in Fig. 2f & 3d-f so that a direct correlation between FCS results and the actual SEM images can be made. The volume in zeptoliter written in the bottom left of each image is deduced from the FCS measurement of the number of molecules inside the nanogap region and the known molecular concentration. The gap size in the bottom right is obtained from the SEM images as the difference between the total length of the rhodium dimer (measured along the main axis) minus the size of each rhodium nanocube (measured along the direction perpendicular to the main axis). For a correlation between the gap sizes deduced from FCS and SEM, please refer to Fig. S10. The nanoantennas are ranked from top left to bottom right as a function of the FCS volume. The alphanumeric code in the top left of each image is our internal reference of each antenna. As the nanocubes forming the antenna are positioned inside the rectangular aperture and are thus below the aluminum surface, they appear dimmer than the other rhodium nanocubes and triangular nanoparticles dispersed on top of the aluminum surface.

## S2. Comparison of gap sizes achieved with other nanofabrication methods

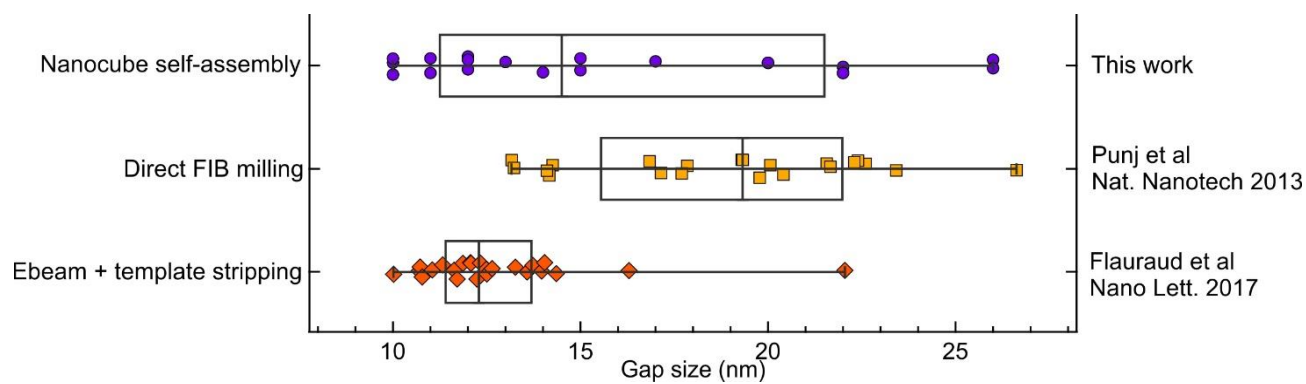

**Figure S2.** Comparison of the estimated gap sizes achieved using different fabrication methods. The markers represent individual nanoantennas while the boxes and whiskers display the median, 25$^{th}$ and 75$^{th}$ percentile and the min/max values. For focused ion beam (FIB) and electron beam lithography (Ebeam) we refer to past works from our group on gold nanogap antennas. This comparison demonstrates that the gap sizes achieved using self-assembly into nanorectangles are quite comparable with other nanofabrication methods. Moreover, let us stress that the main focus of this work is to demonstrate the realization of ultraviolet nanogap antennas and assess their performance for the detection of label-free proteins. We are not aiming at developing a new nanofabrication method.

## S3. Overview of several nanoantennas

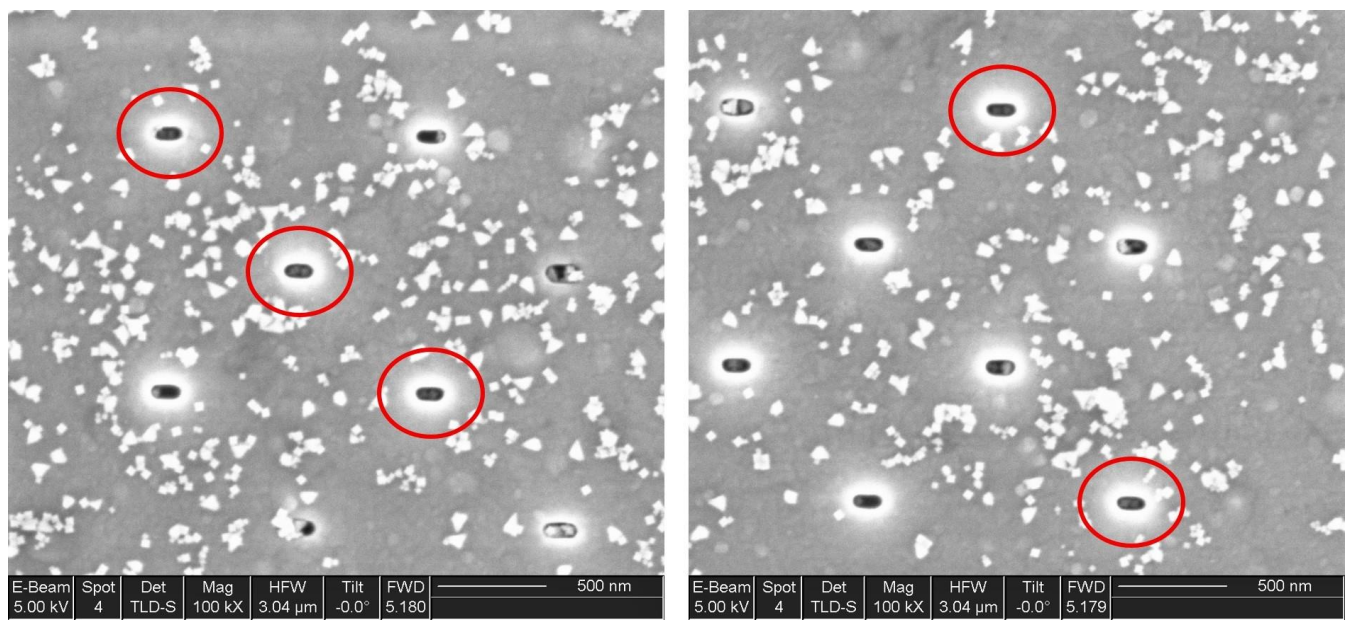

**Figure S3.** SEM images overview of two zones featuring 8 rectangular nanoapertures milled into the aluminum film. The apertures circled in red have two rhodium nanocubes assembled into a nanogap antenna, which are selected for the UV fluorescence experiments. The other apertures are discarded. By doing correlative measurements between the electron microscope and the UV microscope, we can select the self-assembled dimer nanogap antennas. After rinsing, the nanoantennas can be reused several times.

**S4. Electric field intensity maps at 266 nm and at 350 nm**

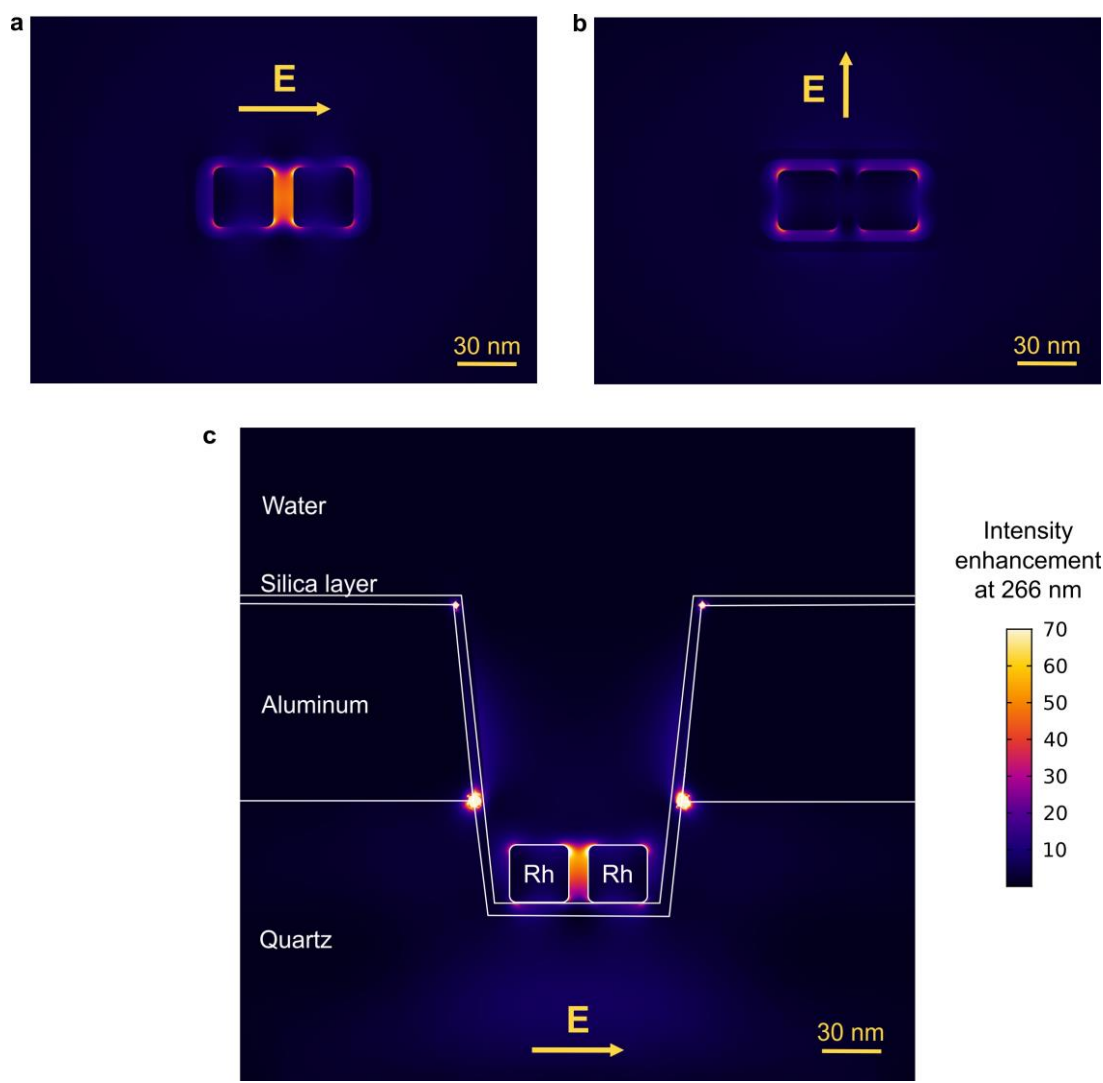

**Figure S4.** 3D numerical simulations of the electric field intensity enhancement at 266 nm for a UV nanogap antenna made of two 30 nm rhodium cubes separated by a 10 nm gap inserted in a rectangular aperture milled into an aluminum film. The arrows indicate the orientation of the incident electric field. The maps a,b are taken 5 nm below the top surface of the nanocubes. All the maps share the same colorscale.

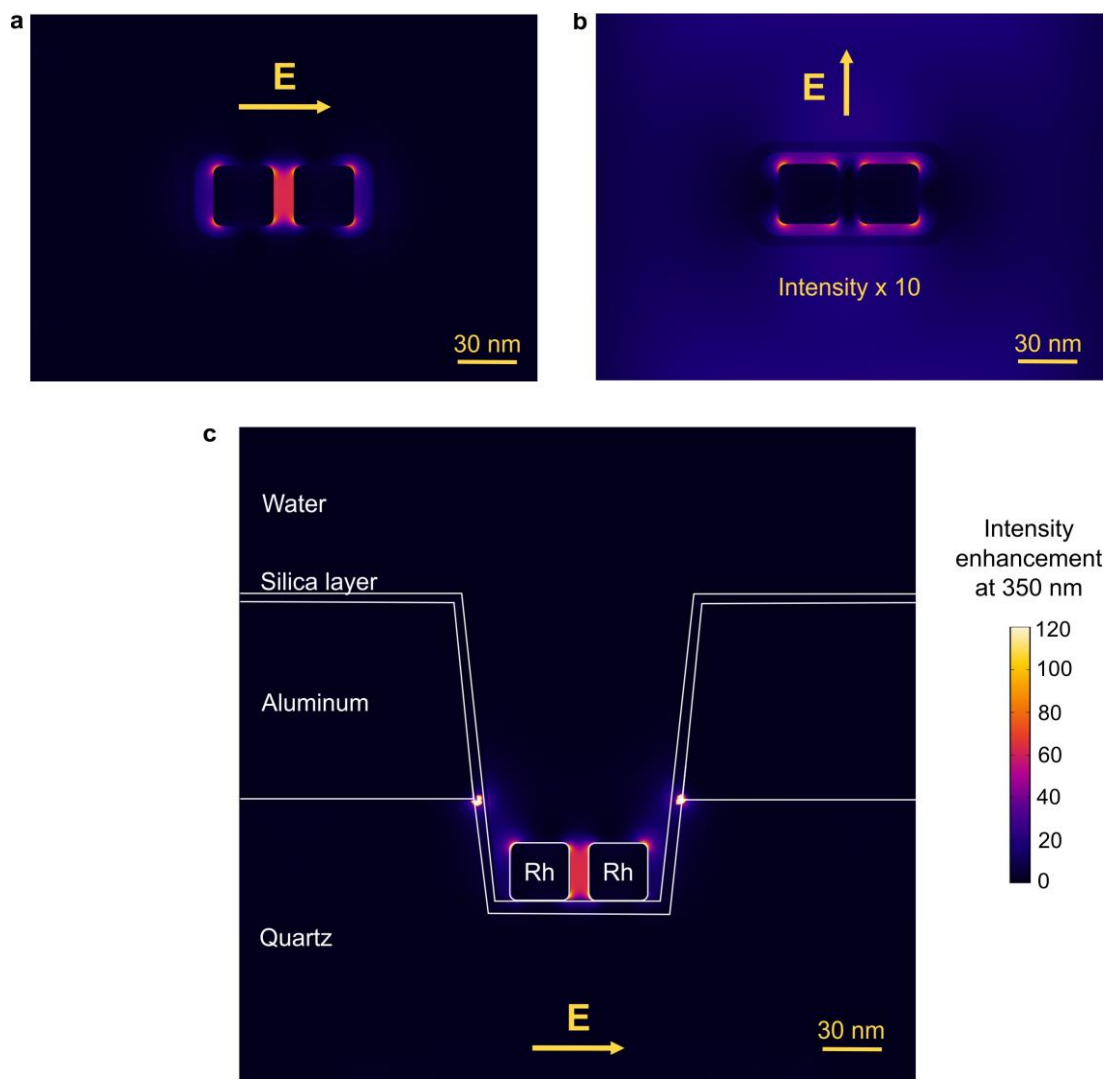

**Figure S5.** 3D numerical simulations of the electric field intensity enhancement at 350 nm for a UV nanogap antenna made of two 30 nm rhodium cubes separated by a 10 nm gap inserted in a rectangular aperture milled into an aluminum film. The arrows indicate the orientation of the incident electric field. The maps a,b are taken 5 nm below the top surface of the nanocubes. So that all the maps share the same colorscale, we have multiplied the intensity for the perpendicular orientation (b) by a constant value of 10x.

## S5. Influence of the nanocube tilt

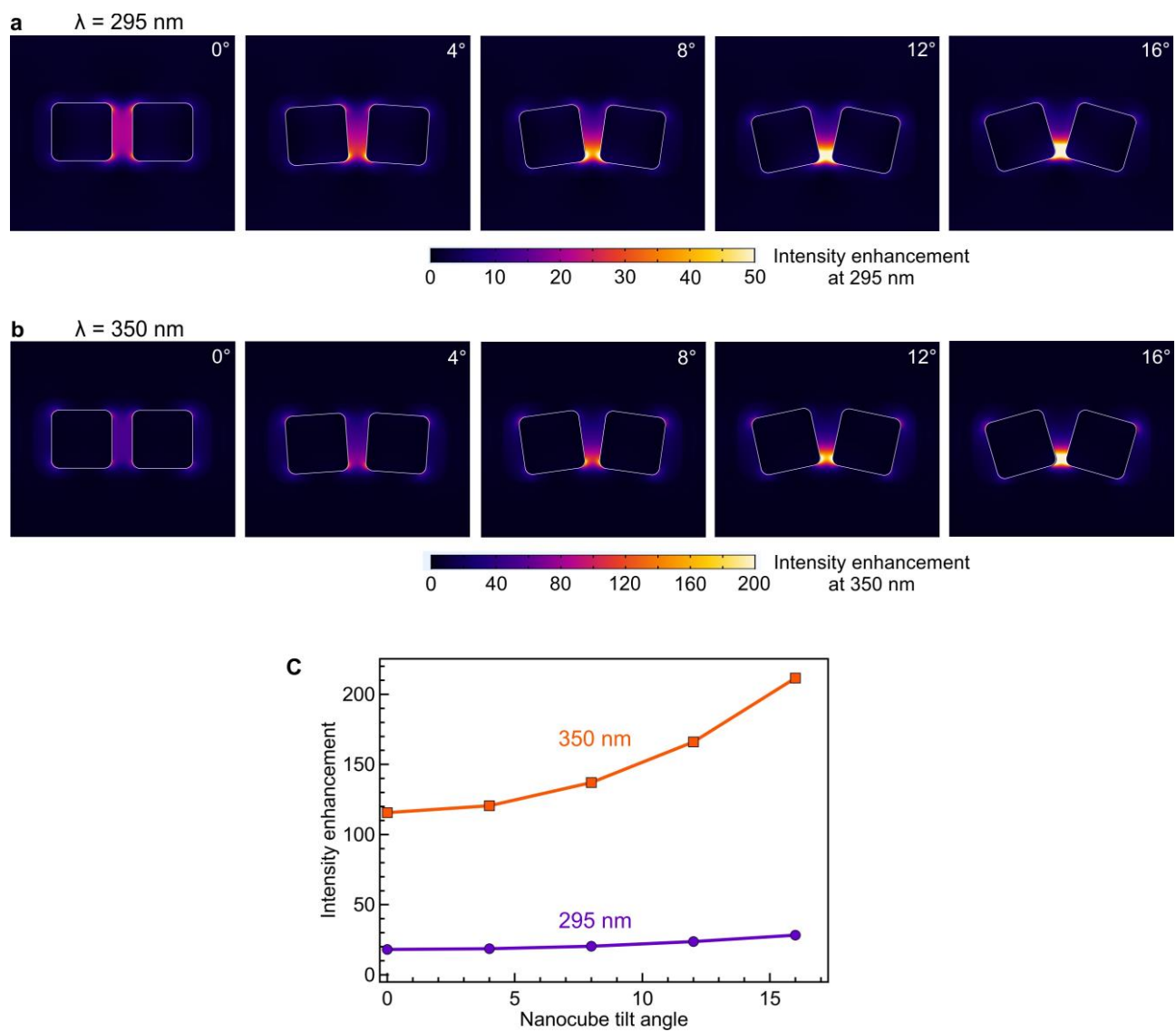

**Figure S6.** Influence of non-perfectly aligned nanocubes on the nanoantenna optical performance. (a) Spatial distributions of the intensity enhancement at 295 nm for different tilt angle of the nanocubes. The angle indicated in the top right corner of each image correspond to the rotation angle of each nanocube in the horizontal XY plane. Each nanocube is rotated by this angle with opposite rotation directions. The nanocube size is 30 nm and their center-to-center distance is kept constant at 40 nm, which corresponds to a 10 nm gap for perfectly aligned nanocubes. (b) Same as (a) for 350 nm wavelength. (c) Spatially-averaged intensity enhancement along a 30 nm vertical line in the middle of the nanogap between the nanocubes.

## S6. Spectral and size dependence of the intensity enhancement for different gap sizes

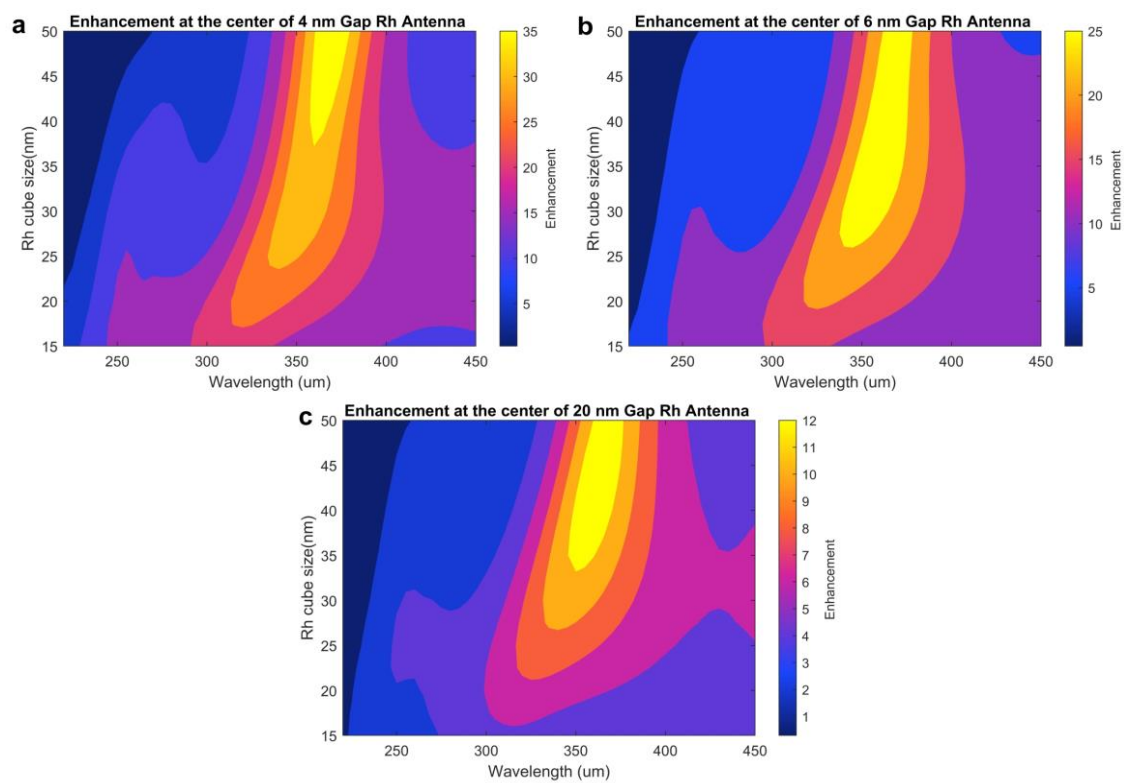

**Figure S7.** Numerical simulations of the spectral dependence of the intensity enhancement in the center of the nanogap antenna as a function of the rhodium nanocube size and gap size. For (a-c) the gap sizes are respectively set to 4, 6 and 20 nm. To speed up the numerical calculations and provide design guidelines, the simulations consider a pair of rhodium nanocubes on a quartz coverslip immersed in water without aluminum layer, so the maximum enhancement values are lower by typically a factor ~4x as compared to the full 3D simulations including the aluminum nanorectangle (Fig. S4 and S5). Zero-mode waveguides and nanorectangle apertures are known to locally enhance the electromagnetic field intensity.[1]

**S7. Autofluorescence emission spectra of proteins used in this work**

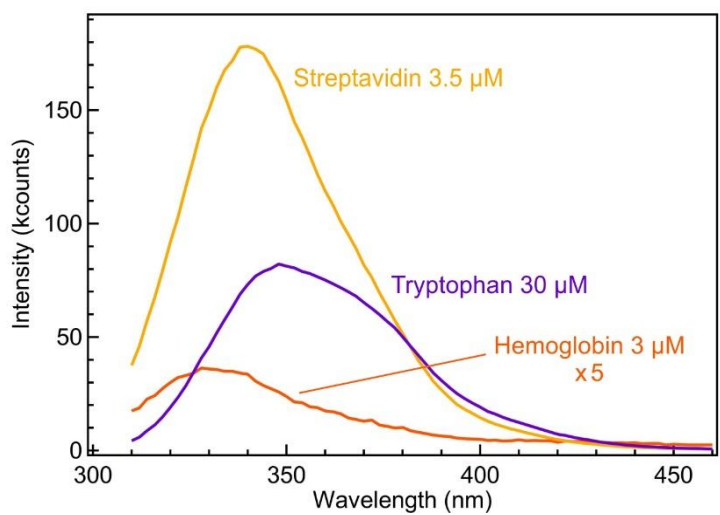

**Figure S8.** Autofluorescence emission spectra of streptavidin and hemoglobin solutions recorded along with a solution of tryptophan diluted in water to serve as a quantum yield reference. The spectra were recorded on a Tecan Spark 10M spectrofluorometer with 260 nm excitation and identical fluorescence detection conditions. The intensity for the hemoglobin spectrum has been multiplied by 5 times to ease viewing on the same graph. To estimate the average quantum yield of the protein autofluorescence, we compute the ratios of the fluorescence intensities integrated over the 310-410 nm spectral region, normalized by the absorbance of the same solutions measured at 260 nm, and we use the calibrated 12% quantum yield of tryptophan in water 100 mM phosphate buffer solution.[2]

## S8. Control FCS in the absence of rhodium nanoantenna

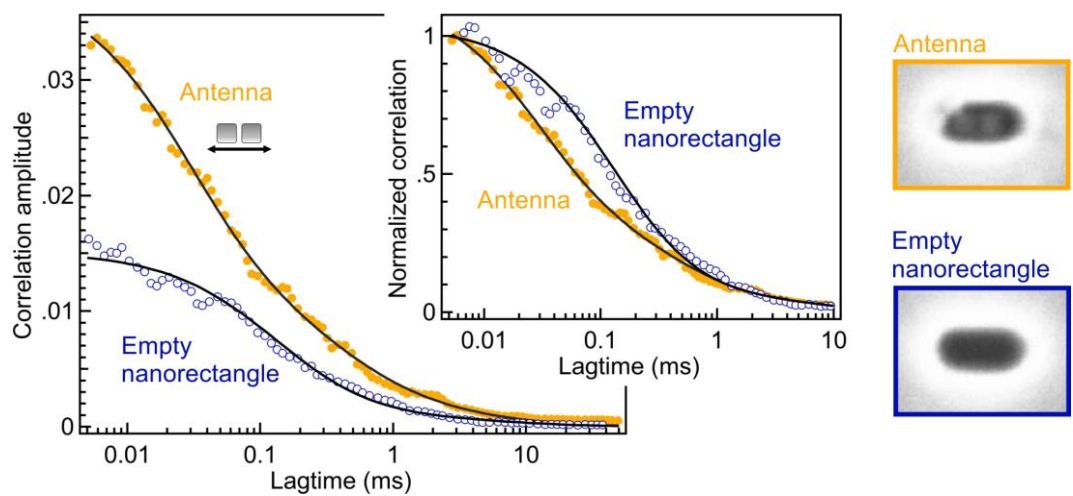

**Figure S9.** Comparison between the FCS correlation functions obtained for a dimer nanoantenna and a nanorectangular aperture without rhodium nanocubes. The conditions are identical to the ones used for Fig. 2 with streptavidin proteins. Dots are experimental data, lines are numerical fits. The insert graph shows the amplitude-normalized FCS functions to highlight the shorter diffusion time in the case of the nanogap antenna. For the experiments on the empty nanorectangle, we recorded an average total intensity $F$ of 8030 counts/s and a background $B$ of 2200 counts/s. From the FCS fit amplitude $\rho_1$ of 0.014, we deduce a number of molecules in the nanorectangle of $N = \left(1 - \frac{B}{F}\right)^2 \frac{1}{\rho_1}$ of 36.8 molecules with a brightness $(F - B)/N$ of 158 counts/s, which is enhanced by 7.2× above the reference 22 counts/s found for streptavidin on our confocal setup.

## S9. Control FCS with a single rhodium nanocube

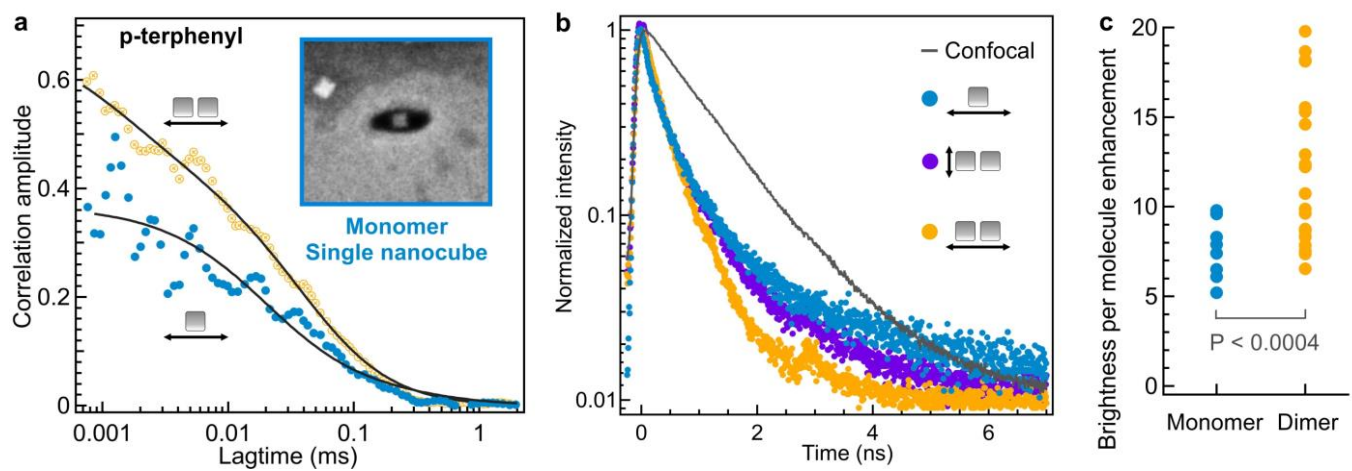

**Figure S10.** Control FCS experiment with a single rhodium nanocube. The conditions are identical to the ones used for Fig. 3 for p-terphenyl dissolved in a glycerol:ethanol mixture (60:40 volume ratio) to increase the viscosity, slow down the diffusion time and facilitate the FCS measurement. The concentration is 10 µM, the 266 nm excitation power is 40 µW. (a) FCS correlation functions. The average total intensity $F$ is 4570 counts/s and the background $B$ is 1700 counts/s. From the FCS fit amplitude $\rho_1$ of 0.34, we deduce a number of molecules of $N = \left(1 - \frac{B}{F}\right)^2 \frac{1}{\rho_1}$ of 1.2 molecules with a brightness $(F - B)/N$ of 2400 counts/s, which is enhanced by 5.2× above the reference 475 counts/s found for p-terphenyl on our confocal setup. (b) Comparison of the normalized time-resolved decay traces, superposing to the data in Fig. 3 the result for the single rhodium nanocube. A slightly longer fluorescence lifetime is observed in the case of the single rhodium cube, which may be related to a reduced quenching rate with a single nanocube instead of two. (c) Comparison of the fluorescence enhancement factors for the brightness per emitter for the single nanocube (monomer) and the dimer of nanocubes, both with parallel excitation. A statistical T-test has been performed to compare the distributions, the resulting P value is written in the graph. The null hypothesis is clearly rejected.

## S10. Correlation between FCS volume and gap size

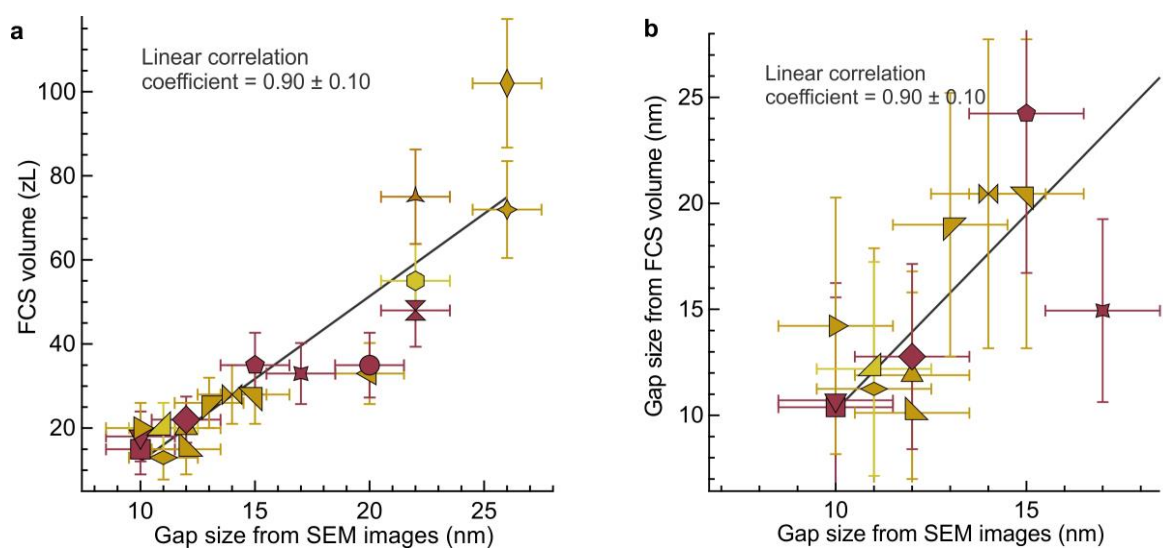

**Figure S11.** Correlation between the gap size obtained from the SEM images with the volume measured with FCS (a) or the gap size deduced from the FCS volume (b) for a selection of the antennas with the smallest gaps. The SEM gap size is the same as the one indicated on Fig. S1. It is obtained from the SEM images as the difference between the total length of the rhodium dimer (measured along the main axis) minus the size of each rhodium nanocube (measured along the direction perpendicular to the main axis). The FCS volume (a) is derived from the FCS measurement of the number of molecules inside the nanogap region and the known molecular concentration. To estimate the gap size from the FCS volume (vertical axis in b), we divide the FCS volume by the lateral area of the rhodium nanocube (square of the average nanocube size measured from the FCS images along the perpendicular direction plus a constant 6 nm to account for the expansion of the detection volume beyond the geometrical limits of the nanocube). Linear fits and Pearson correlation coefficient are indicated on each graph. These results demonstrate the correlation between the FCS results and the SEM gap sizes.

**S11. Brightness per molecule enhancement as a function of gap size from SEM images**

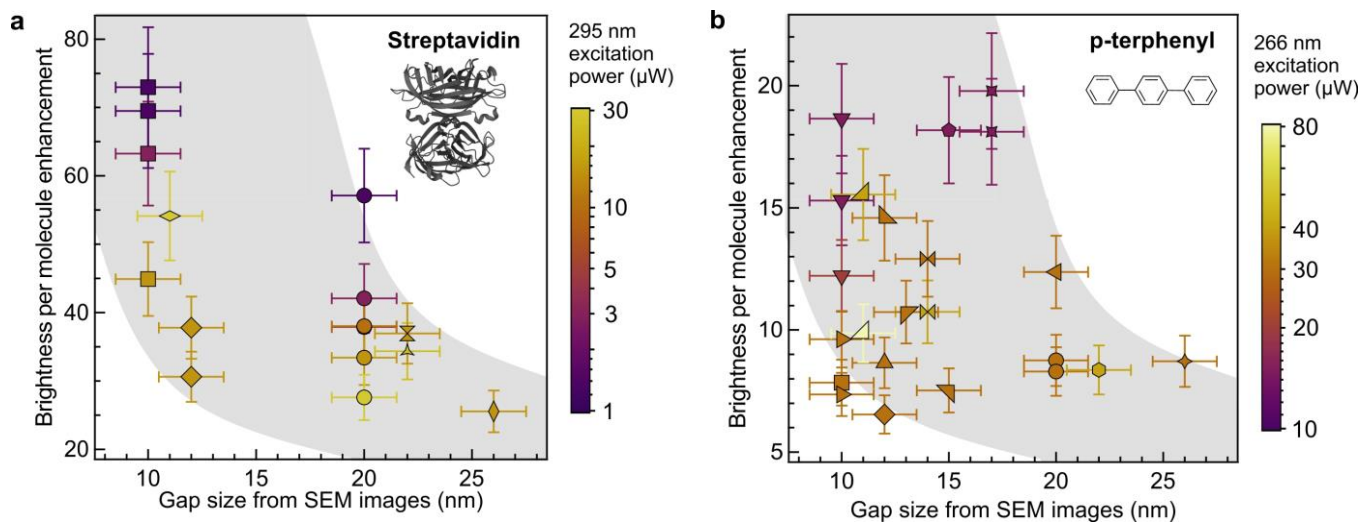

**Figure S12.** Scatter plot of the fluorescence brightness enhancement for streptavidin (a) and p-terphenyl (b) as a function of the gap size deduced from the SEM images in Fig. S1. The enhancement values are the same as in Fig. 2f and 3d, yet the x-axis variable is now the gap size deduced from the SEM images instead of the number of molecules in the nanogap measured with FCS. The different markers indicate the different nanoantennas (same code as in Fig. S1), and the color indicates the excitation power used. The shaded areas are guides to the eyes.

## S12. Nano-antenna enhanced autofluorescence of hemoglobin

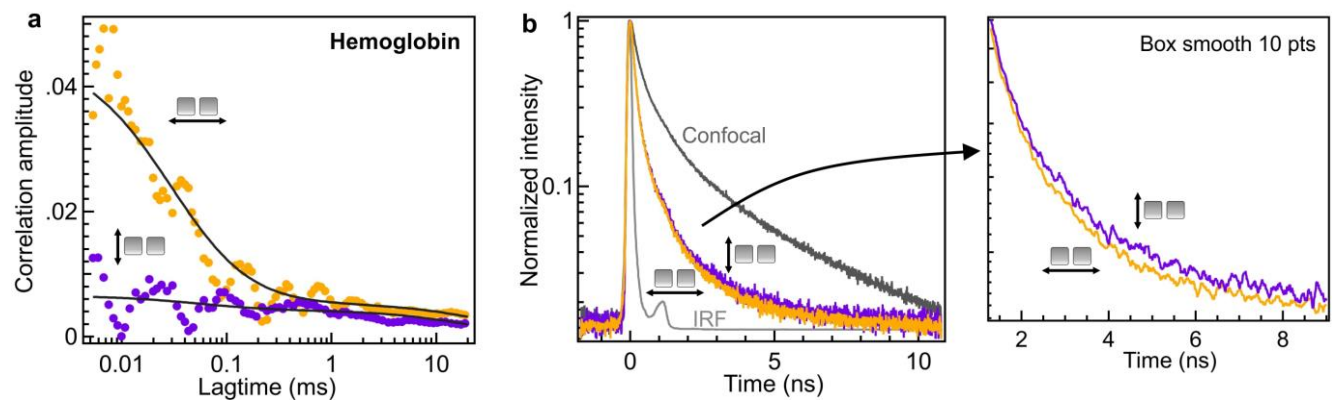

**Figure S13.** FCS and TCSPC data for label-free hemoglobin in a nanogap antenna. The antenna used for this data has the reference number R5s1p5e2-6 for which a 20 nm gap was deduced from the SEM image (Fig. S1). (a) FCS correlation functions, dots are experimental data, lines are numerical fits. The antenna is covered with a 50 µM solution of diffusing label-free hemoglobin proteins. The 295 nm excitation power used here is 15 µW. (b) Normalized time-resolved decay traces corresponding to the data in (a) and to the confocal reference (dark gray). IRF stands for the instrument response function. The insert graph on the right is a close-up view of the decays for excitation polarization set parallel and perpendicular to the dimer's main axis. The data have been smoothed with 10 points box averaging.

## S13. Fitting parameters results

**Table S1.** Fitting parameters for the FCS data.

| Protein | STREPTAVIDIN | | P-TERPHENYL | | HEMOGLOBIN | |
|---|---|---|---|---|---|---|
| Concentration | 50 µM | | 10 µM | | 50 µM | |
| Power (µW) | 15 | | 40 | | 15 | |
| Exc. Polarization | Parallel | Perpendicular | Parallel | Perpendicular | Parallel | Perpendicular |
| F (counts/s) | 5740 | 4860 | 3370 | 2100 | 4070 | 2970 |
| B (counts/s) | 2100 | 2100 | 1700 | 1450 | 2200 | 2200 |
| $\rho_1$ | 0.0276 | -- | -- | -- | 0.028 | -- |
| $\rho_2$ | 0.0093 | 0.0145 | 0.635 | 0.119 | 0.012 | 0.0024 |
| $\rho_3$ | 0.0014 | -- | 0.034 | 0.027 | 0.0053 | 0.0042 |
| $\tau_1$ (ms) | 0.042 | -- | -- | -- | 0.04 | -- |
| $\tau_2$ (ms) | 0.70 | 0.73 | 0.005 | 0.0066 | 0.07 | 0.08 |
| $\tau_3$ (ms) | 8 | -- | 1.4 | 4.6 | 49 | 30 |
| G(0) = $\rho_1 + \rho_2$ | 0.0369 | 0.014 | 0.635 | 0.119 | 0.040 | 0.0024 |
| $N_0$ | 22.2 | -- | 0.8 | -- | 28 | -- |
| $Q_0$ (counts/s) | 124 | -- | 807 | -- | 27.5 | -- |
| N* | 0.90 | 22.2 | 0.16 | 0.8 | 1.9 | 28 |
| Q* (counts/s) | 987 | 124 | 6530 | 807 | 583 | 27.5 |
| $Q_{ref}$ (counts/s) | 22 | 22 | 475 | 475 | 3.5 | 3.5 |
| Brightness enhancement | 44.9 | 5.6 | 13.75 | 1.7 | 166 | 7.9 |

**Table S2.** Fitting parameters for the TCPSC data. For the average lifetime of p-terphenyl in parallel case, we took only the first and second component into consideration to compute the intensity-averaged lifetime.

| Protein | STREPTAVIDIN | | P-TERPHENYL | | HEMOGLOBIN | |
|---|---|---|---|---|---|---|
| Exc. Polarization | Parallel | Perpendicular | Parallel | Perpendicular | Parallel | Perpendicular |
| $\tau_1$ (ns) | 0.025 | | 0.025 | | 0.025 | |
| $\tau_2$ (ns) | 0.166 | | 0.140 | | 0.243 | |
| $\tau_3$ (ns) | 0.860 | | 0.450 | | 1.740 | |
| $I_1$ (%) | 19.74 | 23.91 | 7.48 | 30.85 | 11.09 | 18.52 |
| $I_2$ (%) | 32.61 | 23.91 | 25.99 | -- | 60.27 | 52.15 |
| $I_3$ (%) | 47.64 | 52.18 | 66.53 | 69.15 | 28.64 | 29.33 |
| $\tau_{average\ intensity}$ (ns) | 0.47 | 0.49 | 0.114 | 0.32 | 0.65 | 0.64 |

## S14. Decay rates enhancement with rhodium nanogap antennas

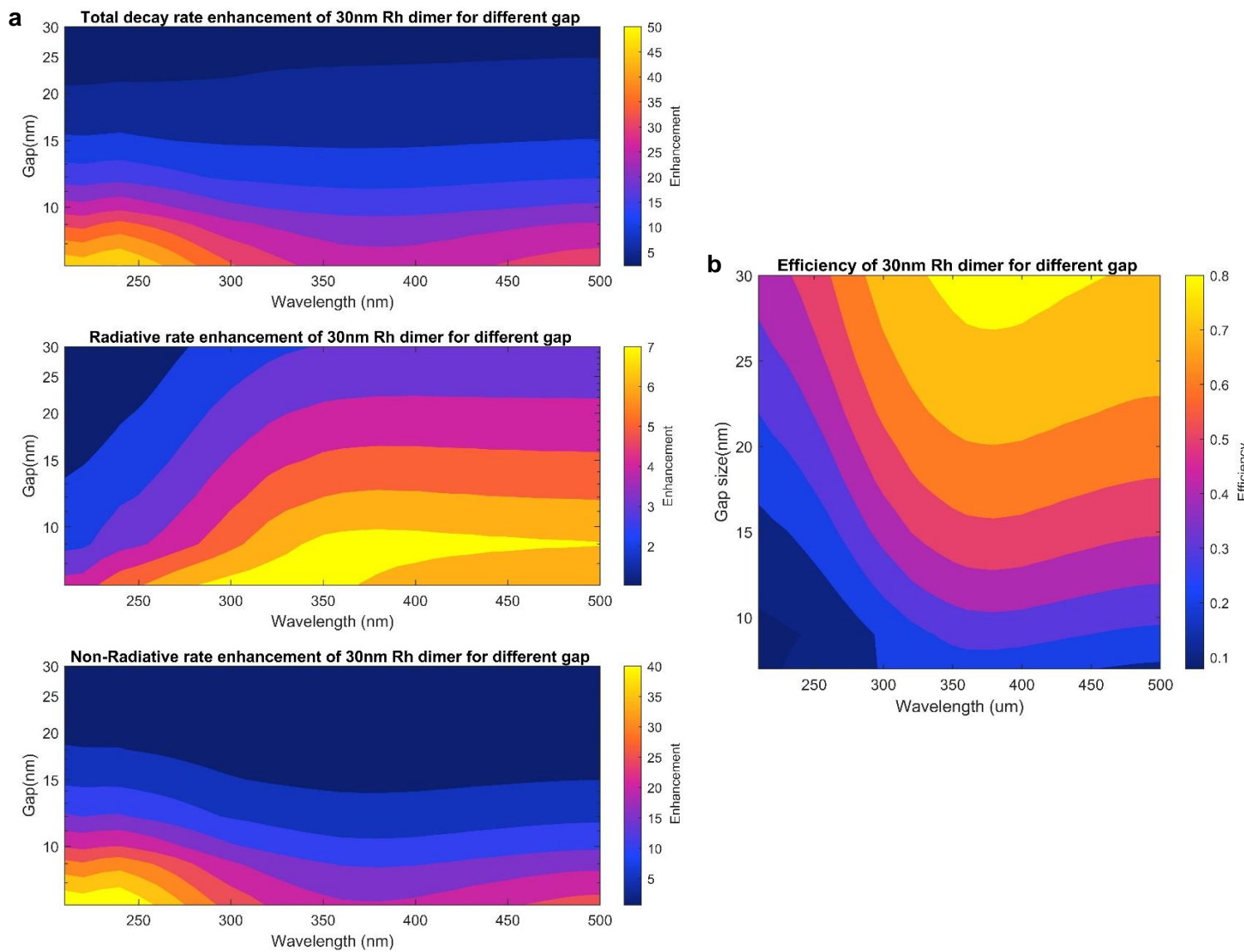

**Figure S14.** (a) Numerical simulations of the total, radiative and nonradiative decay rate constant enhancement for a point dipole located in the center of a rhodium dimer nanogap antenna oriented along the antenna's main axis as a function of the emission wavelength and the gap size. The cube size is constant at 30 nm. All rates are normalized to the dipole's radiative rate in free space. (b) Simulations of the antenna radiative efficiency (ratio of radiative rate to total decay rate) as a function of the emission wavelength and the gap size for a perfect point dipole emitter with parallel orientation located in the center of the nanogap. The rhodium cube size is constant at 30 nm.

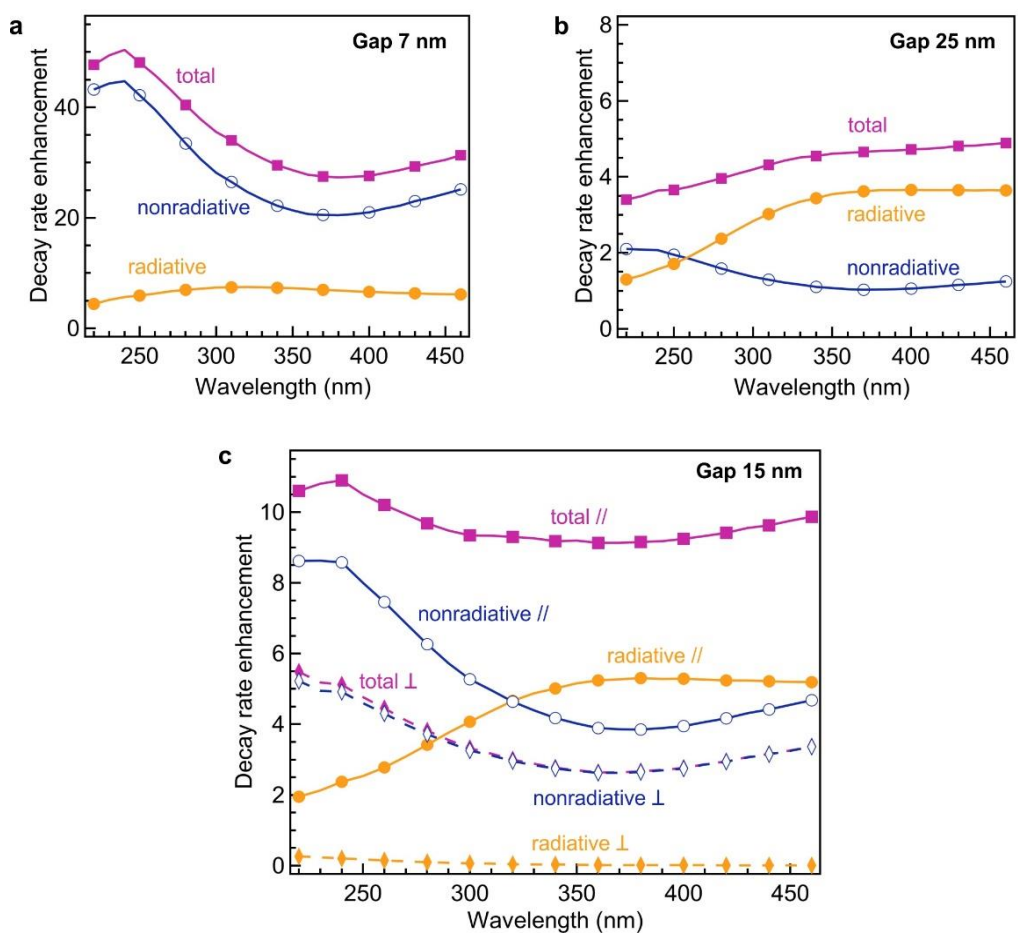

**Figure S15.** (a-c) Numerical simulations of the enhancement of the decay rate constants as a function of the emission wavelength for a perfect point dipole emitter with parallel orientation located in the center of the nanogap. In (c), the enhancement factors for a dipole with perpendicular orientation are displayed with dashed lines and diamond markers. The rhodium cube size is constant at 30 nm. The gap size is 7 nm in (a), 25 nm in (b) and 15 nm in (c). All rates are normalized respective to the dipole radiative rate in free space.

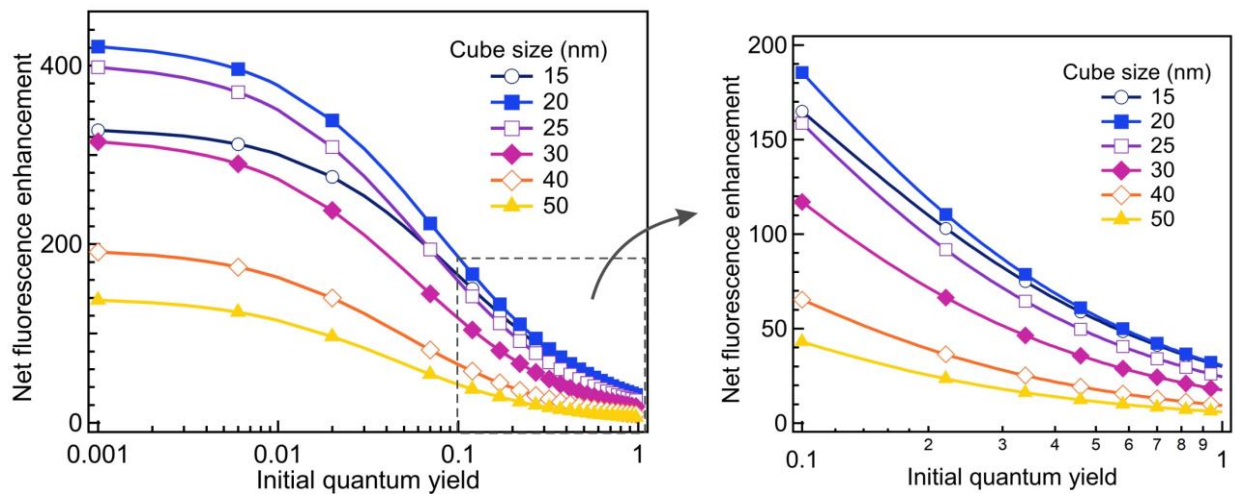

**Figure S16.** Simulations of the maximum fluorescence brightness enhancement as a function of the initial quantum yield for a point dipole located in the center of the nanogap with an orientation parallel to the dimer's main axis. The gap size is kept constant at 10 nm. The excitation wavelength is 295 nm and the emission is 350 nm. The right panel is a close-up view of the zone with quantum yields between 0.1 and 1.

## S15. Experimental determination of the photokinetic rates in the rhodium nanoantenna

To determine experimentally the influence of the rhodium nanoantenna on the photokinetic rates, we use the following approach and notations: for the confocal reference, the total decay rate constant $\Gamma_{tot}^0 = \Gamma_{rad}^0 + \Gamma_{nr}^0$ is the sum of the radiative $\Gamma_{rad}^0$ and nonradiative $\Gamma_{nr}^0$ decay rate constants. It also amounts to the inverse of the fluorescence lifetime $1/\tau_0$. The quantum yield is $\phi_0 = \Gamma_{rad}^0/\Gamma_{tot}^0$. The different values are summarized in Table S3 for the various molecules used here.

In presence of the nanoantenna, the lifetime is shortened, and becomes $1/\tau^* = \Gamma_{rad}^* + \Gamma_{nr}^0 + \Gamma_{loss}^*$. Here we consider that the radiative decay rate constant $\Gamma_r^*$ is enhanced (Purcell effect), that the internal nonradiative decay rate constant $\Gamma_{nr}^0$ is unaffected by the photonic environment and that an additional nonradiative decay channel $\Gamma_{loss}^*$ is introduced to account for the extra losses into the free electron cloud in the metallic antenna.[3,4] The fluorescence brightness enhancement $\eta_F$ corresponds to the product of the gains in excitation intensity $\eta_{exc}$, quantum yield $\eta_\phi$, and collection efficiency $\eta_{coll}$.[5] The quantum yield gain can be further written as the ratio between the gains in the radiative rate $\eta_{\Gamma rad} = \Gamma_{rad}^*/\Gamma_{rad}^0$ and the total decay rate $\eta_{\Gamma tot} = \Gamma_{tot}^*/\Gamma_{tot}^0$, so the fluorescence enhancement becomes $\eta_F = \eta_{exc}\,\eta_{coll}\,\eta_{\Gamma rad}/\eta_{\Gamma tot}$ (here $\eta_{\Gamma tot}$ is also equivalent to the reduction in the fluorescence lifetime). Using these notations, the fluorescence enhancement can be rewritten to clearly show the dependence with the initial quantum yield $\phi_0$:[6–8]

$$\eta_F = \eta_{exc}\,\eta_{coll}\,\eta_{\Gamma rad}\,\frac{1}{1 - \phi_0 + \phi_0\,(\Gamma_{rad}^* + \Gamma_{loss}^*)/\Gamma_{rad}^0}$$

With the 0.8 numerical aperture of our microscope objective, the laser beam can still be considered to be moderately focused (this numerical aperture corresponds to a maximum angle of 34° in quartz). In this case, the reciprocity theorem states that the gain in excitation intensity amounts to the products of the gains in collection efficiency times the gain in radiative rate: $\eta_{exc} = \eta_{coll}\,\eta_{\Gamma rad}$.[9] We also use recent numerical simulations of the collection efficiency gain to estimate its value to $\eta_{coll} = 1.44$ for all the different molecules here.[10]

From the measurements of the brightness enhancement $\eta_F$ together with the fluorescence lifetimes in confocal and in the nanoantenna and the knowledge of the quantum yield $\phi_0$ in homogeneous solution, we can compute back all the different rate constants, including the losses to the metal $\Gamma_{loss}^*$. The main results are summarized in Table S3. Interestingly, despite more than two orders of magnitude difference in the initial quantum yields of p-terphenyl, streptavidin and hemoglobin, we find consistent results in the excitation gain, radiative gain and loss decay rate, indicating a common electromagnetic origin for these effects. The orientation-averaged and position-averaged excitation gain $\eta_{exc} = 15.5 \pm 3.8$ appears in correct agreement with the numerical simulations Figs. 1c-f, as well as the radiative gain $\eta_{\Gamma rad} = 10.8 \pm 2.6$. The loss rate constant $\Gamma_{loss}^* = 1.25 \pm 0.3$ ns$^{-1}$ appears to be also a preserved feature among our different experiments.

**Table S3.** Photokinetic parameters. The rate constants are expressed in ns$^{-1}$, the lifetimes are in ns.

| | $\phi_0$ | $\tau_0$ | $\Gamma_{tot}^0$ | $\Gamma_{rad}^0$ | $\Gamma_{nr}^0$ | $\eta_F$ | $\tau^*$ | $\eta_{exc}$ | $\eta_{\Gamma rad}$ | $\Gamma_{loss}^*$ | $\phi^*$ | $\eta_\phi$ |
|---|---|---|---|---|---|---|---|---|---|---|---|---|
| P-terphenyl | 0.93 | 1.00 | 1.00 | 0.93 | 0.07 | 16.1 | 0.11 | 12.4 | 8.6 | 1.5 | 0.84 | 0.9 |
| Streptavidin | 0.035 | 1.50 | 0.67 | 0.02 | 0.64 | 65 | 0.47 | 14.4 | 10.0 | 1.3 | 0.11 | 3.1 |
| Hemoglobin | 0.005 | 2.10 | 0.48 | 0.00 | 0.47 | 120 | 0.65 | 19.7 | 13.6 | 1.0 | 0.021 | 4.2 |

## S16. Comparison of the enhancement factors with aluminum nanogap antennas

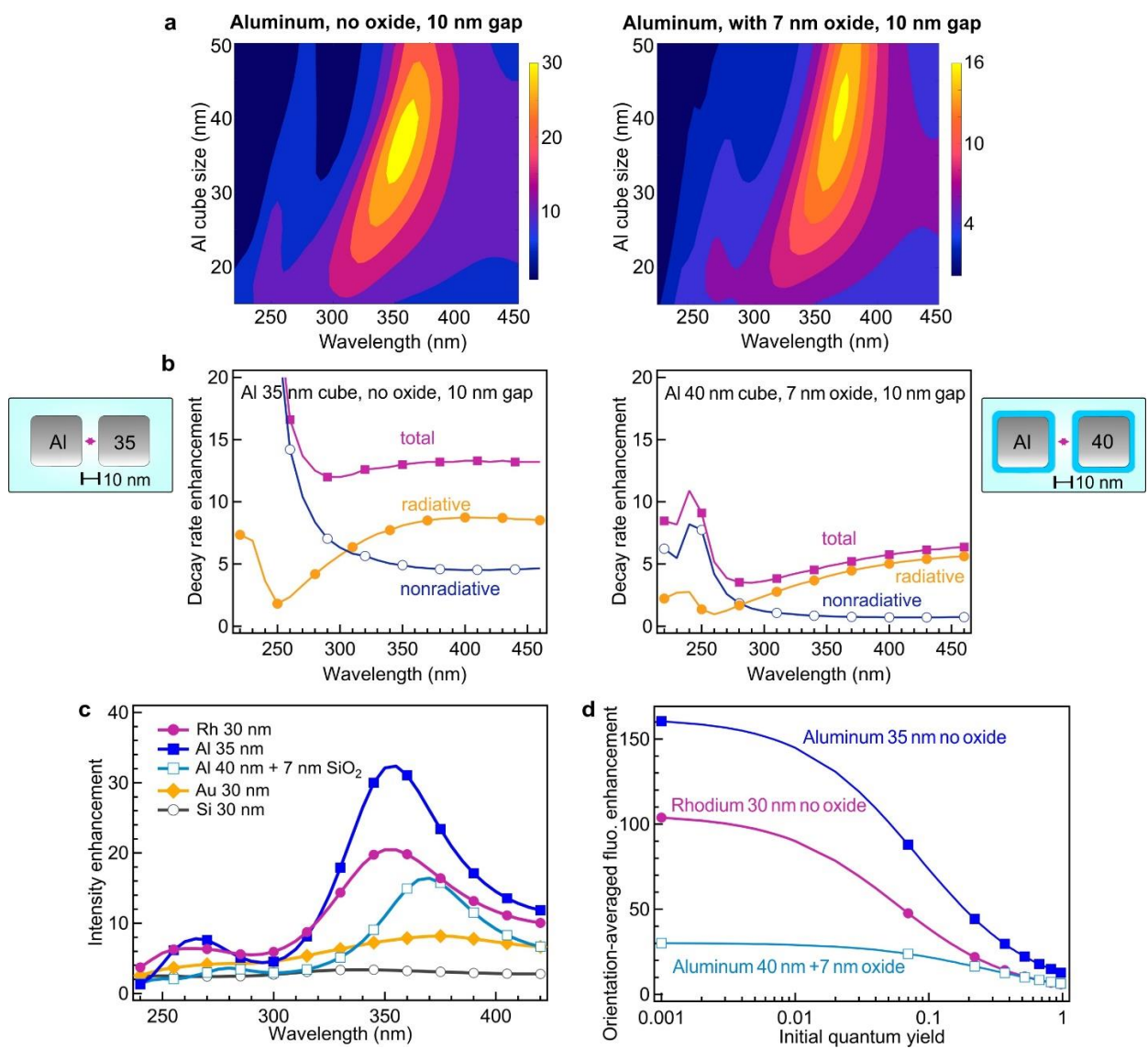

**Figure S17.** Comparison of rhodium and aluminum nanocube antennas in water. (a) Spectral dependence of the intensity enhancement in the center of the nanogap as a function of the nanocube size. The gap is constant at 10 nm. The left image is for pure aluminum without any oxide layer. On the right image, we have added a 7 nm thick conformal silica layer surrounding the nanocube to simulate the influence of an extra corrosion protection layer. Corrosion of aluminum in water environment is a major issue, especially under UV illumination.[11–14] Based on the results in (a), we select a nanocube size of 35 nm for pure aluminum, and a 40 nm size in the presence of silica. (b) Decay rate enhancement for a perfect dipole emitter with parallel orientation located in the center of the 10 nm nanogap. All rates are normalized respective to the dipole radiative rate in free space. (c) Comparison of the intensity enhancement in the center of the 10 nm gap between nanocubes made of different materials. The nanocube sizes and the materials are indicated in the figure legend. We have considered amorphous silicon. (d) Comparison of the net fluorescence brightness enhancement for three different materials as a function of the emitter's initial quantum yield in free space. The gap size is kept constant at 10 nm. The excitation wavelength is 295 nm for rhodium and 266 nm for aluminum since this wavelength gives a slightly better overall enhancement. The emission wavelength is 350 nm and is averaged over the three orientation directions.

## S17. Protein information and sequences

**Table S4.** Information about the proteins used in this work.

| Name | **Streptavidin** | **Hemoglobin** | |
|---|---|---|---|
| Organism | Streptomyces avidinii | Homo Sapiens (Human) | |
| UniPROT reference | P22629 | $\alpha$ subunit: P69905 | $\beta$ subunit: P68871 |
| RCSB PDB structure ref | 2RTR | 6BB5 | |
| Sigma Aldrich product number | S4762 | H7379 | |
| Form | Homotetramer | Heterodimer of $\alpha$ and $\beta$ subunits: $\alpha_2\beta_2$ | |
| Molecular weight of the full protein (Da) | 75,336 | 62,512 | |
| Monomer sequence length (aa) | 183 | $\alpha$ subunit: 142 | $\beta$ subunit: 147 |
| Monomer sequence (tryptophan W and tyrosine Y highlighted) | MRKIVVAAIAVSLTTVSITASASADPSKDSKAQVSAAEAGITGT**W****Y**NQLGSTFIVTAGADGALTGT**Y**ESAVGNAESR**Y**VLTGR**Y**DSAPATDGSGTALG**W**TVA**W**KNN**Y**RNAHSATT**W**SGQ**Y**VGGAEARINTQ**W**LLTSGTTEANA**W**KSTLVGHDTFTKVKPSAASIDAAKKAGVNNGNPLDAVQQ | $\alpha$ subunit: MVLSPADKTNVKAA**W**GKVGAHAGE**Y**GAEALERMFLSFPTTKT**Y**FPHFDLSHGSAQVKGHGKKVADALTNAVAHVDDMPNALSALSDLHAHKLRVDPVNFKLLSHCLLVTLAAHLPAEFTPAVHASLDKFLASVSTVLTSK**Y**R | $\beta$ subunit: MVHLTPEEKSAVTAL**W**GKVNVDEVGGEALGRLLVV**Y**P**W**TQRFFESFGDLSTPDAVMGNPKVKAHGKKVLGAFSDGLAHLDNLKGTFATLSELHCDKLHVDPENFRLLGNVLVCVLAHHFGKEFTPPVQAA**Y**QKVVAGVANALAHK**Y**H |
| Tryptophan count per monomer | 6 | 1 | 2 |
| Tyrosine count per monomer | 6 | 3 | 3 |
| Total tryptophan residues per protein | 24 | 6 | |
| Total tyrosine residues per protein | 24 | 12 | |
| Extinction coefficient $\varepsilon$ at 280 nm ($M^{-1}$ $cm^{-1}$) | 169,360 | 57,630 | |